\documentclass[11pt]{article}
\RequirePackage[OT1]{fontenc}
\usepackage{amsthm,amsmath,natbib}
\RequirePackage[colorlinks,citecolor=blue,urlcolor=blue]{hyperref}
\usepackage{amssymb}
\usepackage{graphicx,overpic,amsmath}
\usepackage{bm}
\usepackage{upgreek}
\usepackage{url}
\usepackage{color}
\usepackage[overload]{textcase}
\usepackage[ruled]{algorithm2e}
\usepackage{makecell}
\usepackage{natbib}
\usepackage{multirow}

\usepackage{diagbox}

\newcommand{\pkg}[1]{{\normalfont\fontseries{b}\selectfont #1}}

\let\code=\texttt

\newtheorem{example}{Example}

\newcommand{\ba}{\begin{array}}
\newcommand{\ea}{\end{array}}
\newcommand{\bt}{\begin{tabular}}
\newcommand{\et}{\end{tabular}}
\newcommand{\btb}{\begin{table}}
\newcommand{\etb}{\end{table}}
\newcommand{\bc}{\begin{center}}
\newcommand{\ec}{\end{center}}
\newcommand{\bea}{\begin{eqnarray}}
\newcommand{\eea}{\end{eqnarray}}
\newcommand{\Bea}{\begin{eqnarray*}}
\newcommand{\Eea}{\end{eqnarray*}}
\newcommand{\beq}{\begin{equation}}
\newcommand{\eeq}{\end{equation}}

\DeclareMathOperator*{\argmin}{arg\,min}

\newcommand{\p}{{\rm I}\kern-0.18em{\rm P}}
\newcommand{\1}{{\rm 1}\kern-0.24em{\rm I}}
\newcommand{\E}{{\rm I}\kern-0.18em{\rm E}}
\newcommand{\R}{{\rm I}\kern-0.18em{\rm R}}

\def \bfm#1{\mbox{\boldmath$#1$}}
\def \es2{$E(s^2)$}

\def \R {{\bfm R}}

\def \0{{\bf 0}}

 \def \s2{{\sigma^2}}

\title{Imbalanced classification: a paradigm-based review}

\author{ Yang Feng$^{1}$,
	Min Zhou$^{2}$,
	Xin Tong$^{3}$
	\footnote{ $^{1}$ Department of Biostatistics, School of Global Public Health,  New York University, 715 Broadway, New York, NY 10003, USA. (e-mail: yang.feng@nyu.edu). 
		$^{2}$ Division of Science and Technology, Beijing Normal University-Hong Kong Baptist University United International College, Zhuhai, China. (e-mail: minzhou@uic.edu.cn).
		$^{3}$ Department of Data Sciences and Operations, Marshall School of Business, University of Southern California, Los Angeles, CA 90089, USA. (e-mail: xint@marshall.usc.edu, zhoumin@marshall.usc.edu). This work is partially supported by NSF CAREER Grant DMS-1554804 and NIH Grant R01 GM120507. Feng and Zhou contribute equally to this work. Tong is the corresponding author.
	} 
\\
	\\$^1$ New York University \\
	$^{2}$ BNU-HKBU United International College\\
    $^{3}$ University of Southern California
}

\date{}

\begin{document}
\baselineskip 16pt

\maketitle

\begin{center}
{\begin{abstract}  
A common issue for classification in scientific research and industry is the existence of imbalanced classes. When sample sizes of different classes are imbalanced in training data, naively implementing a classification method often leads to unsatisfactory prediction results on test data. Multiple resampling techniques have been proposed to address the class imbalance issues. Yet, there is no general guidance on when to use each technique. In this article, we provide a paradigm-based review of the common resampling techniques for binary classification under imbalanced class sizes. The paradigms we consider include the classical paradigm that minimizes the overall classification error, the cost-sensitive learning paradigm that minimizes a cost-adjusted weighted type I and type II errors, and the Neyman-Pearson paradigm that minimizes the type II error subject to a type I error constraint. Under each paradigm, we investigate the combination of the resampling techniques and a few state-of-the-art classification methods. For each pair of resampling techniques and classification methods, we use simulation studies and a real data set on credit card fraud to study the performance under different evaluation metrics.  From these extensive numerical experiments, we demonstrate under each classification paradigm, the complex dynamics among resampling techniques, base classification methods, evaluation metrics, and imbalance ratios. We also summarize a few takeaway messages regarding the choices of resampling techniques and base classification methods, which could be helpful for practitioners.

\end{abstract}
}
\end{center}


\section{Introduction}
Classification is a widely studied type of supervised learning problem with extensive applications.  A myriad of classification methods (e.g., logistic regression, support vector machines, random forest, neural networks, boosting), which we refer to as the \textit{base classification methods} in this paper,  have been developed to deal with different distributions of data \citep{kotsiantis2007supervised}. However, in the case where the classes are of different sizes (i.e., the imbalanced classification scenario), naively applying the existing methods could lead to undesirable results. Some prominent applications include defect detection \citep{arnqvist2021efficient}, medical diagnosis \citep{chen2016empirical}, fraud detection \citep{wei2013effective}, spam email filtering \citep{youn2007comparative}, text categorization \citep{zheng2004feature}, oil spills detection in satellite radar images \citep{kubat1998machine}, land use classification \citep{ranneby2011nonparametric}.   To address the class size imbalance scenario, there has been extensive research on developing different methods  \citep{sun2009classification, lopez2013insight, haixiang2017learning}. Some popular tools include resampling techniques \citep{lopez2013insight,alahmari2020comparison,anis2020analysis}, direct methods \citep{lin2002support, ling2004decision, zhou2005training, sun2007cost,  qiao2010weighted}, post-processing methods \citep{castro2013novel}, as well as different combinations of these tools. The most common and understandable class of approaches is resampling techniques. However, there lacks a consensus about when and how to use them.

In this work, we aim to provide some guidelines on using resampling techniques for imbalanced binary classification. We first disentangle the general claims of undesirability in classification results under imbalanced classes, via listing a few common paradigms and evaluation metrics. To decide which resampling technique to use, we need to be clear on the paradigms as well as the preferred evaluation metrics. Sometimes, the chosen paradigm and the evaluation metric are not compatible, which makes the problem unsolvable by any technique. When they are, we will show that the optimal resampling technique depends on both the paradigm and the base classification method.  

There are different degrees of data imbalance.  We characterize this degree by the \emph{imbalance ratio} (IR)   \citep{garcia2012effectiveness}, which is the ratio of the sample size of the majority class and that of the minority class. In real applications, IR can range from $1$ to more than $1,000$. For instance, a rare disease occurs only in 0.1\% of the human population \citep{beaulieu2014forge}. We will show that different IRs might demand different combinations of resampling techniques and base classification methods.

This review conducts extensive simulation experiments as well as a real data set on credit card fraud to concretely illustrate the dynamics among data distributions, IR, base classification methods, and resampling techniques. This is the \emph{first} time that such dynamics are explicitly examined. To the best of our knowledge, this is also the first time that a review paper uses running simulation examples to demonstrate the advantages and disadvantages of the reviewed methods.  Through simulation and real data analysis, we give practitioners a look into the complicated nature of the imbalanced data problem in classification, even if we narrow our search to the resampling techniques only.    For important applications where data distributions can be approximately simulated, practitioners are encouraged to mimic our simulation studies and properly evaluate the combinations of resampling techniques and base classification methods. In the end, we summarize a few takeaway messages regarding the choices of resampling techniques and base classification methods, which could be helpful for practitioners. 

The rest of the review is organized as follows. In Section \ref{sec::class_paradigm}, we describe three classification paradigms and discuss their corresponding objectives. Then, we introduce a matrix of classification algorithms as pairs of resampling techniques and the base classification methods in Section \ref{sec:matrix-class}. Section \ref{sec:eval_metric} provides a list of commonly used evaluation metrics for imbalanced classification. In Sections \ref{sec:simulation} and \ref{sec:real}, we conduct a systematic simulation study and a real data analysis to evaluate the performance of different combinations of resampling techniques and base classification methods, under different paradigms, data distributions, and IRs, in terms of various evaluation metrics. We conclude the review with a short discussion in Section \ref{sec:discussion}.

\section{Three Classification Paradigms}\label{sec::class_paradigm}
In this section, we review three classification paradigms that are defined by different objective functions. Concretely, we consider the Classical Classification (CC) paradigm that minimizes the overall classification error (Section \ref{subsec:CC}), the Cost-Sensitive (CS) learning paradigm that minimizes the cost-adjusted weighted type I and type II errors (Section \ref{subsec:CS}),  and the Neyman-Pearson (NP) paradigm that minimizes the type II error subject to a type I error constraint (Section \ref{subsec:NP}).

Assume  $X\in\mathcal{X}\subset \R^d$ is a random  vector of $d$ features, and  $Y\in \{0, 1\}$ is  the  class label. Let $\p(Y=0)=\pi_0$ and $\p(Y=1)=\pi_1=1-\pi_0$.  Throughout the article, we label the minority class as   0 and the majority class as  1 (i.e.,  $\pi_0\le\pi_1$). Also, for language consistency,  we call  class 0  the negative class and  class 1  the positive class. Please note that the minority class might be referred  to as   ``positive'' in  medical applications.

\subsection{Classical Classification paradigm}\label{subsec:CC}
A classifier  is defined as $\phi: \mathcal{X}\rightarrow \{0,1\}$, which is a mapping from the feature space to the label space.  The overall classification error (risk) is naturally defined as 
$R(\phi)=\E[\1(\phi(X)\neq Y)]=\p(\phi(X)\neq Y)$, where $\1(\cdot)$ is the indicator function. In binary classification, most existing classification methods focus on the minimization of the overall classification error (risk)  \citep{hastie2009elements,james2013introduction}. In this article, this paradigm is referred to as \emph{Classical Classification (CC) Paradigm.} Under this paradigm, the \emph{CC oracle}   $\phi^*$ is a classifier that minimizes the population risk; that is, 
$$\phi^*=\argmin_{\phi}R(\phi)\,.$$
It is well known that $\phi^*=\1(\eta(x)>1/2)$, where $\eta(x)=\E(Y|X=x)$ is the regression function \citep{koltchinskii2011oracle}. In practice, we construct a classifier $\hat \phi$ based on finite sample $\{(X_i, Y_i), i=1,\cdots,n\}$ using some classification method.

Popular the CC paradigm is, it may not be the ideal choice when the class sizes are imbalanced.  By the \emph{law of total probability}, we decompose the overall classification error as a weighted sum of type I and  II errors, that is, 
$$R(\phi)=\pi_0 R_0(\phi)+\pi_1R_1(\phi)\,,$$
where $R_0(\phi)=\p(\phi(X)\neq Y|Y=0)$ denotes the (population) type I error (the conditional probability of misclassifying a class 0 observation as class 1);  and $R_1(\phi)=\p(\phi(X)\neq Y|Y=1)$ denotes the (population) type II error (the conditional probability of misclassifying a class 1 observation as class 0).   
However, in many practical applications, we may want to treat  type I and II errors differently  under two common scenarios.  One is the \emph{asymmetric error importance} scenario. In this scenario, making one type of error (e.g., type I error) is more serious than making the other type of error (e.g., type II error).  For instance, in severe disease diagnosis,  misclassifying a diseased patient as healthy could lead to missing the optimal treatment window while misclassifying a healthy patient as diseased can lead to patient anxiety and incur additional medical costs. The other is the \emph{imbalanced class proportion} scenario.  Under this scenario, $\pi_0$ is much smaller than $\pi_1$, and minimizing the overall classification error could sometimes result in a larger type I error.  For applications that fit these two scenarios, the overall classification error may not be the optimal choice to serve the users' purpose, either as an optimization criterion or as an evaluation metric. Next, we will introduce two other paradigms that have been used the address the asymmetric error importance and imbalanced class proportion issues.

\subsection{Cost-Sensitive learning paradigm}\label{subsec:CS}
In the asymmetric error importance and imbalanced class proportion scenarios introduced at the end of Section \ref{subsec:CC}, 
the cost of type I error is usually higher than that of type II error. For example, in spam email filtering, the cost of misclassifying a regular email as spam is much higher than the cost of misclassifying spam as a regular email.  A popular approach to incorporate different costs for these two types of errors is the \emph{Cost-Sensitive (CS) learning} paradigm \citep{elkan2001foundations, zadrozny2003cost}. Let $C(\phi(X), Y)$ being the cost function for classifier $\phi$ at observation pair $(X,Y)$. Let $C_0=C(1,0)$ and $C_1=C(0,1)$ being the costs of type I and II errors, respectively. For the correct classification result, we have $C(0,0)=C(1,1)=0$. Then,  CS learning minimizes the  expected misclassification cost \citep{kuhn2013applied}:
\begin{align*}
R_c(\phi) &= \E C(\phi(X), Y)\\
&=C_0\p(\phi(X)=1, Y=0)+C_1\p(\phi(X)=0, Y=1)\\
&=C_0\p(\phi(X)=1|Y=0)\p(Y=0) + C_1\p(\phi(X)=0|Y=1)\p(Y=1)\\
&=C_0\pi_0R_0(\phi)+C_1\pi_1R_1(\phi)\,.
\end{align*}

There are  primarily two types of approaches in the literature on CS learning. The first type is called \emph{direct methods}, which builds a cost-sensitive learning classifier by incorporating the different misclassification costs into the training process of the base classification method. For instance, there has been much work on CS decision tree \citep{ling2004decision, bradford1998pruning, turney1994cost}, CS boosting  \citep{sun2007cost,wang2010boosting, lopez2015cost}, CS SVM \citep{qiao2009adaptive},  and CS neural network \citep{zhou2005training}. The second type is usually referred to as \emph{postprocessing methods}, in such a way that we adjust the decision threshold with the base classification algorithm unmodified. An example of this can be found in \cite{domingos1999metacost}. Some additional references on cost-sensitive learning include \cite{lopez2012analysis,lopez2013insight,haixiang2017learning,voigt2014threshold,zhang2016imbalanced,zou2016finding}.

In this review, we focus on the \emph{postprocessing methods} as it combines well with any existing base classification algorithm without the need to change its internal mechanism, which is also better understood among practitioners. In addition, it serves the purpose of making an informative comparison among different learning paradigms across different classification methods. On the population level, 
with the knowledge of $C_0$ and $C_1$, the \emph{CS oracle}  is 
$$\phi_c^*=\argmin_{\phi}R_{c}(\phi)=\1\left(\eta(x)>\frac{C_0}{C_0+C_1}\right),$$
which reduces to the CC oracle $\phi^*$ when $C_0=C_1$. 

Although CS learning has its merits on the control of asymmetric errors,  its drawback is also apparent because it is sometimes difficult or immoral to assign the values of costs $C_0$ and $C_1$. In most applications, including the severe disease classification, these costs are unknown and cannot be easily provided by experts. One way to extricate from this dilemma is to set the majority class misclassification cost $C_1=1$ and  the minority class misclassification cost $C_0=\pi_1/\pi_0$ \citep{castro2013novel}.

\subsection{Neyman-Pearson paradigm}\label{subsec:NP}
Besides requiring the knowledge of costs for different misclassification errors, the CS learning paradigm does not provide an explicit probabilistic control on type I error under a pre-specified level. Even if the practitioner tunes the empirical type I error equal to the pre-specified level, the population-level type I error still has a non-trivial chance of exceeding this level \citep*{tong2016survey, tong2018neyman}.  To deal with this issue,  another emerging statistical framework to control asymmetric error is  called \emph{Neyman-Pearson (NP)  paradigm} \citep{cannon2002learning,rigollet2011neyman,tong2013plug,tong2016survey,tong2018neyman}, which aims to minimize type II error $R_1(\phi)$ while controlling  type I error  $R_0(\phi)$ under a desirable level. The corresponding \emph{NP oracle} is 
$$\phi_\alpha^*=\argmin_{\phi: R_0(\phi)\leq\alpha}R_1(\phi)\,,$$
where $\alpha$ is a targeted upper bound for type I error. It can be shown that $\phi^*_{\alpha}(\cdot) = \1(\eta(\cdot) > D^*_{\alpha})$ for some properly chosen $D^*_{\alpha}$.  Unlike $1/2$ or $C_0/(C_0+C_1)$, $D^*_{\alpha}$ is not known unless one has access to the distribution information.      \cite{tong2018neyman} proposed an umbrella algorithm for NP classification, which adapts existing scoring-type classification methods (e.g., logistic regression, support vector machines, random forest) by choosing an order-statistics based thresholding level so that the resulting classifier has type I error bounded from above by $\alpha$ with high probability. This thresholding mechanism, along with the thresholds $1/2$ and $C_0/(C_0+C_1)$ for CC and CS paradigms respectively, will be systematically studied in combination with several state-of-the-art base classification methods in numerical studies.

\subsection{A summary of three classification paradigms}
For readers' convenience, we summarize the three classification paradigms with their corresponding objectives and oracle classifiers in Table \ref{cp}.

\begin{table}[!htbp]
	\caption{ Three types of classification paradigms in binary classification.\label{cp}}
	\centering
	\begin{tabular}{lll}
		\hline\hline
		Paradigm     &       Objective &Oracle Classfier   \\
		\hline
		Classical              &    Minimize the overall classification error &$\phi^*=\argmin_{\phi}R(\phi)$                  \\
		Cost-Sensitive          &  Minimize the expected misclassification cost  &$\phi_c^*=\argmin_{\phi}R_{c}(\phi)$                      \\
		\multirow{2}{*}{Neyman-Pearson}        &   Minimize type II error while controlling     &\multirow{2}{*}{$\phi_\alpha^*=\argmin\limits_{\phi: R_0(\phi)\leq\alpha}R_1(\phi)$ }  \\
		& type I error under  $\alpha$ & \\
		\hline
	\end{tabular}
	
\end{table}

 \section{A Matrix of Algorithms for Imbalanced Classification} \label{sec:matrix-class}
In this section, we introduce a matrix of algorithms for imbalanced classification, which consists of combinations of resampling techniques  and base classification methods. 

To fix idea, assume among the $n$ observation pairs $\{(X_i, Y_i), i=1,\cdots,n\}$, there are $n_0$ observations with $Y_i=0$ (the minority class) and $n_1$ observations with $Y_i=1$ (the majority class). Then, the \emph{imbalance ratio}  IR = $n_1/n_0$.

\subsection{Resampling techniques}\label{subsec::resampling}
To address the imbalanced classification problem under one of the three classification paradigms described in Section \ref{sec::class_paradigm},  resampling techniques are often used to create a new training dataset by balancing the number of data points in the minority and majority classes in order to alleviate the effect of class size imbalance in the process of classification.  \cite{lopez2013insight} pointed out that about one-third of their reviewed papers have used resampling techniques. They are usually divided into three categories:  undersampling, oversampling, and hybrid methods.

The undersampling methods directly discard a subset of observations of the majority class. It includes two main versions: the \emph{cluster-based undersampling} and \emph{random undersampling} \citep{yen2009cluster,kumar2014undersampled,sun2015novel,haixiang2017learning}.   In the \emph{cluster-based undersampling},  a clustering algorithm is applied to cluster the majority class such that the number of clusters is equal to that of the data points in the minority class (i.e., $n_0$ clusters), and then one point is randomly selected from each cluster. Nevertheless, the clustering process could be quite slow when $n_1$ is large.  \emph{Random undersampling} is a simpler and more efficient approach, which randomly eliminates the data points from the majority class to make it of size $n_0$. By undersampling, the processed training data set is a combination of $n_0$ randomly chosen data points from the  majority class and all ($n_0$) data points from the minority class. However, undersampling may lead to loss of information as a large portion of the data from the majority class is discarded.

 The oversampling method, on the other hand, increases the number of data points in the minority class from $n_0$ to $n_1$ while keeping the observations from the majority class intact. The leading two approaches are \emph{random oversampling} and \emph{SMOTE} \citep{han2005borderline,he2008adasyn,garcia2012surrounding,beaulieu2014forge,nekooeimehr2016adaptive}. \emph{Random oversampling}, as a counterpart of \emph{random undersampling}, is perhaps the most straightforward approach to duplicate the data points of minority class randomly.  One version of the approach  samples $n_1-n_0$ observations with replacement from the minority class and add them to the new training set.  The approach \emph{SMOTE} is the acronym for the ``\emph{Synthetic Minority Over-sampling Technique}" proposed by  \cite{chawla2002smote}. It generates $n_1-n_0$ new synthetic data points for the minority class by interpolating pairs of  $k$ nearest neighbors.  We review the details of SMOTE in Algorithm \ref{algo:smote}. Compared with undersampling, oversampling methods usually require longer training time and could cause over-fitting. A popular extension of SMOTE is the Borderline-SMOTE (BLSMOTE) \citep{han2005borderline}, which only oversamples the minority observations near the borderline and the essential step to generate the data point is similar to the SMOTE algorithm in Algorithm \ref{algo:smote} (see \cite{han2005borderline} for a detailed description for BLSMOTE).\\
\begin{algorithm}[H]\caption{ SMOTE  \citep{chawla2002smote}\label{algo:smote}} 
	For any data point of minority class $X_i=(X_{i1},X_{i2},\ldots,X_{id})^\top$, the multiple $N=IR-1$, number of nearest neighbors $K$\\
	{\bf Step 1:} Find the $K$ nearest neighbor points of $X_i$ in the minority class: $X_{i_1},\ldots,X_{i_k}$\;
	{\bf Step 2:}
	\For{$j=1:N$}
	{   randomly choose one of the $K$ nearest neighbor points: $X_{i_j}=(X_{i_j1},X_{i_j2},\ldots,X_{i_jd})^\top$\;
		generate a random number $r_s\sim Unif[0,1]$\;
		generate the synthetic data point for the minority class as $X_j^*=(X_{j1}^*,X_{j2}^*,\ldots,X_{jd}^*)^\top$, where $X_{js}^*=X_{is}+r_s*(X_{i_js}-X_{is})$,  $s=1,\ldots,d$.
	}
	{\bf Return} $X_1^*,\ldots,X_N^*$ as new synthetic data points.
\end{algorithm}

The hybrid method is just a combination of undersampling and oversampling methods \citep{cao2014ensemble,cateni2014method,diez2015random,saez2015smote}. It simultaneously decreases the number of data points from the majority class and increases the number of data points from the minority class  to $n_h$, where the above described undersampling and oversampling methods can be used. The hybrid method could serve as an option that balances the goodness of fit, computational cost as well as  robustness of the classifier.

\subsection{Classification methods}\label{subsec:base-classifier}
Using any of the resampling methods, we will arrive at a new training dataset that has balanced classes. Naturally, we can apply any existing base classification method on this new dataset coupled with one of the paradigms described in Section \ref{sec::class_paradigm}. 

Many classification methods have been extensively studied. The well-known ones include   decision trees (DT) \citep{safavian1991survey}, $k$-nearest neighbors (KNN) \citep{altman1992introduction},  Linear discriminant analysis (LDA) \citep{mclachlan2004discriminant},  logistic regression (LR) \citep{nelder1972generalized},  na\"ive bayes (NB) \citep{rish2001empirical}, neural network (NN) \citep{rumelhart1985learning}, random forest (RF) \citep{breiman2001random}, support vector machine (SVM) \citep{cortes1995support}, and XGBoost (XGB) \citep{chen2016xgboost}, among others.

To learn more about these methods, we refer the readers to a review of classification  methods  \citep{kotsiantis2007supervised}  and a book on statistical learning \citep{hastie2009elements}.

\subsection{A summary of the matrix of algorithms}\label{subsec:matrix-algo}
\begin{figure}[!htbp]
	\centering
	\includegraphics[width=4.8 in]{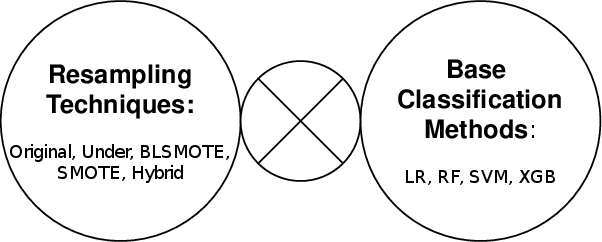}\\
	\caption{A summary of the matrix of algorithms.}\label{algorithm matrix}
\end{figure}

In numerical studies, we consider a matrix of classification algorithms shown in Figure \ref{algorithm matrix}, as combinations of resampling techniques described in Section \ref{subsec::resampling} and four (out of many) state-of-the-art classification methods described in Section \ref{subsec:base-classifier}.

In Figure \ref{algorithm matrix},   ``Original" refers to \emph{no resampling}, ``Under" refers to \emph{random undersampling} and ``Hybrid" refers to \emph{a hybrid of random undersampling and SMOTE}. Note here we chose random undersampling, SMOTE, and BLSMOTE as representatives of undersampling and oversampling methods due to their popularity among practitioners.  The readers can easily study other types of resampling technique and classification method combinations by adapting the companion code from this review. 

\begin{figure}[!htbp]
	\centering
	\includegraphics[width=5 in]{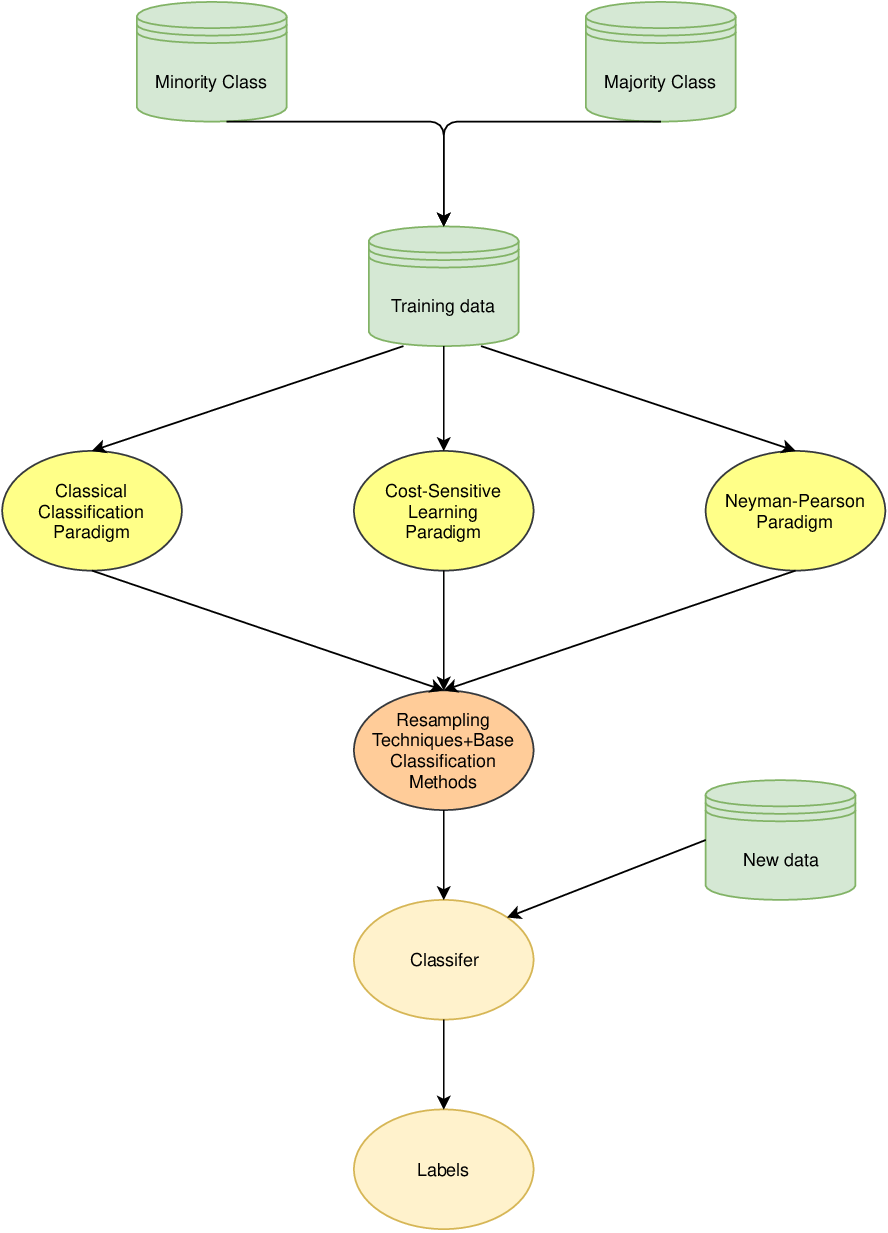}\\
	\caption{A flow chart for imbalanced classification with a paradigm-oriented view.}\label{flow}
\end{figure}

In the numerical studies, we will conduct a comparative study on those $20$ combinations described in Figure \ref{algorithm matrix} under each of the three paradigms introduced in Section \ref{sec::class_paradigm} with the IR varying from $1$ to $128$, in terms of different evaluation metrics which will be introduced in the next section. A flowchart demonstrating our imbalanced classification system can be found in Figure \ref{flow}.

\section{Evaluation Metrics}\label{sec:eval_metric}

In this section, we will review several popular evaluation metrics to compare the performance of different classification algorithms. 

For a given classifier, suppose that it classifies the $i$-th observation $X_i$ to $\hat Y_i$  ($Y_i$ denotes the true label). Then, the classification results can be summarized into the four terms: \emph{True Positives} $\mbox{TP} = \sum_{i=1}^n \1(Y_i=1, \hat Y_i=1)$, \emph{False Positives} $\mbox{FP}=\sum_{i=1}^n\1(Y_i=0, \hat Y_i=1)$, \emph{False Negatives} $\mbox{FN}=\sum_{i=1}^n\1(Y_i=1, \hat Y_i=0)$, and \emph{True Negatives} $\mbox{TN} = \sum_{i=1}^n \1(Y_i=0, \hat Y_i=0)$.  These four terms are usually summarized in the so-called \emph{confusion matrix} (Table \ref{tb:confusion}).

\begin{table}[!htbp]
	\caption{ Confusion matrix for a two-class problem.}\label{tb:confusion}
	\centering
	\begin{tabular}{ccccccccc}
		\hline\hline
		&     Predicted Class 0 &  Predicted Class 1  \\
		True Class 0               &       TN                   &    FP\\
		True Class 1             &       FN                     &    TP \\
		\hline
	\end{tabular}
\end{table}

Note that in Table \ref{tb:confusion}, the class 0 is being regarded as the ``negative class". In practice, sometimes we may need to set class 0 as the ``positive class". 

Then, the \emph{empirical risk} can be denoted as
$$\hat{R}=\hat{\pi}_0\hat{R}_0+\hat{\pi}_1\hat{R}_1=\frac{FP+FN}{TP+FP+TN+FN}\,,$$
where $\hat{\pi}_0=(TN+FP)/(TP+FP+TN+FN)$, $\hat{\pi}_1=(FN+TP)/(TP+FP+TN+FN)$   are the empirical proportions of Class 0 and 1;  $\hat{R}_0$ and $\hat{R}_1$  are the empirical type I  and II errors, respectively, that is,   
$$\hat{R}_0=\frac{FP}{TN+FP},\  \ \hat{R}_1=\frac{FN}{FN+TP}\,.$$
Similarly, for given costs $C_0$ and $C_1$, the \emph{empirical misclassification cost} is expressed as
$$\hat{R}_c=C_0\hat{\pi}_0\hat{R}_0+C_1\hat{\pi}_1\hat{R}_1\,.$$

Another popular synthetic metric in the imbalanced classification literature is the $F$-score (also $F_1$-score or $F$-measure,  \citep{bradley1997use}) for class 0, which is the harmonic mean of Precision  and Recall:
where $Precision_0=TN/(TN+FN)$ and $Recall_0=TN/(TN+FP)$.
Similarly, we can also define $F$-score for class 1 as
$$F\mbox{-score (class 1)}=\frac{2}{Precision_1^{-1}+Recall_1^{-1}}\,,$$
where $Precision_1=TP/(TP+FP)$ and $Recall_1=TP/(TP+FN)$.  Here, we set $F\mbox{-score (class 0)}$ or $F\mbox{-score (class 1)}$ to 0 if
the corresponding precision or recall is undefined or equal to 0. 

When the parameter in a classification method (e.g., the threshold of scoring functions) is varied, we usually get different trade-offs between type I and type II errors. A popular tool to visualize these trade-offs is the Receiver Operating Characteristic (ROC) curve \citep{bradley1997use,huang2005using}. The area under the ROC curve (ROC-AUC) provides an aggregated  measure for the method's performance. ROC-AUC has been used extensively to compare  the performance of different classification methods. However, when the data is highly imbalanced, the ROC curves can present an overly optimistic view of classifiers' performance  \citep{davis2006relationship}. Precision-Recall (PR) curves and their AUCs (PR-AUC) have been advocated as an alternative metric  when dealing with imbalanced data \citep{goadrich2004learning,singla2005discriminative}. Note that we also have two versions of PR-AUC, depending on which class we call ``positive": PR-AUC (class 0) and PR-AUC (class 1).

Now, we summarize  all of the metrics discussed in Table \ref{metric}. 

\begin{table}[!htbp]
	\caption{ Various evaluation metrics.}\label{metric}
	\centering
	\begin{tabular}{ccccccccc}
		\hline\hline
		Metric &     Formula &     \\
		\hline
		Risk                                           &       $(FP+FN)/(TP+FP+TN+FN)  $                \\
		Type I  error($\hat{R}_0$)           &      $FP/(TN+FP) $           \\
		Type II error($\hat{R}_1$)          & $FN/(FN+TP)$   \\
		Cost                     &  $C_0\hat{\pi}_0\hat{R}_0+C_1\hat{\pi}_1\hat{R}_1$    \\
		$F$-score  (class 0)       &  $2/(Precision_0^{-1}+Recall_0^{-1})$\\
		$F$-score  (class 1)       &  $2/(Precision_1^{-1}+Recall_1^{-1})$\\
		ROC-AUC                  & The area under the ROC curve\\
		PR-AUC (class 0)            &  The area under the PR  curve when class 0 is negative\\
		PR-AUC (class 1)            &  The area under the PR  curve when class 0 is positive\\
		\hline
	\end{tabular}
\end{table}

\section{Simulation}\label{sec:simulation}
In this section, we conduct extensive simulation studies to compare the matrix of 20 combinations of classification methods and resampling approaches introduced in Section \ref{sec:matrix-class} under each of the three classification paradigms described in Section \ref{sec::class_paradigm} when the IR varies, using evaluation metrics reviewed in Section \ref{sec:eval_metric}.

\subsection{Data generation process\label{subsec:dgp}}
We consider the following two examples with different data generation mechanisms. 
\begin{example}\label{ex:1}The conditional distributions for each class are multivariate $t_4$ distributions with a common covariance matrix but different mean vectors. Concretely, 
	$$\text{Class 0}: X|(Y=0)\sim t_4\left(\mu^0,\Sigma\right),\ \ \ \text{Class 1}: X|(Y=1)\sim t_4\left(\mu^1,\Sigma\right),$$
	where $\mu^0=(0, 0, 0, 0, 0)^\top$, $\mu^1=(2, 2, 2, 0, 0)^\top$,  and 
	$$\Sigma=\left(\begin{matrix}
	1&0.5&0.25&0&0\\
	0.5&1&0.5&0&0\\
	0.25&0.5&1&0&0\\
	0&0&0&1&0\\
	0&0&0&0&1\\
	\end{matrix}\right)\,.$$
	\begin{enumerate}
	 \item [(a)] To have a precise control on the imbalance ratio (IR), we explicitly generate $n_0=300$ observations from the minority class (class 0) and $n_1$ observations from the majority class, where $\mbox{IR} = n_1/n_0$ is a pre-specified value varying in $\{2^i, i=0,1,\cdots,7\}$. This leads to a training sample $\{(X_i, Y_i), i=1,\cdots,n\}$ where $n=n_0+n_1$. Following the same mechanism, we also generate a test sample with size $m$ consisting of $m_0=2000$ and $m_1=m_0\times IR$ observations from class 0 and 1, respectively. This generation mechanism guarantees the same IR for both training and test samples. 
		\item [(b)] To observe the influence of different IR for test samples,  we fix  $\mbox{IR}_{train}=8$ for training samples and vary $IR_{test}$ in $\{2^i, i=0,1,\cdots,7\}$ for test  samples. The parameters $n_0$ and $m_0$ are 300 and 2000 respectively; and $n_1=300\times 8=2400$, $m_1=m_0\times IR_{test}$.
	\end{enumerate}
\end{example}

\begin{example}\label{ex:2}
	The conditional distributions for each class are multivariate Gaussian vs. a mixture of multivariate Gaussian. Concretely, 
	\begin{align}
	\mbox{Class } 0: \quad & X|(Y=1)\sim \mathcal{N}\left(\frac{1}{2}(\mu^0+\mu^1),\Sigma\right),\\
	\mbox{Class } 1: \quad & X|(Y=0)\sim \frac{1}{2}\mathcal{N}\left(\mu^0,\Sigma\right) + \frac{1}{2}\mathcal{N}\left(\mu^1,\Sigma\right),
	\end{align}
	where $\mu^0$, $\mu^1$ and $\Sigma$ are the same as Example \ref{ex:1}. The remaining data generation mechanism is the same as in Example \ref{ex:1}. As a result, we also have Example 2(a) with the same training and testing IR and 2(b) where we fix the training IR and vary the testing IR. 
\end{example}

\subsection{Implementation details}\label{subsec:imple_details}
Regarding the resampling methods, we consider the following four options. 
\begin{itemize}
	\item No resampling (Original): we use the training dataset as it is without any modification.
	\item    Random undersampling (Under):  we keep all the $n_0$ observations in the minority class and randomly sample $n_0$ observations without replacement from the majority class. Then, we have a balanced data set in which each class is of size $n_0$. 
	\item  Oversampling (SMOTE, BLSMOTE): we keep all the $n_1$ observations in the majority class. We use SMOTE and BLSMOTE (R Package \pkg{smotefamily}, v1.3.1, \citealt{smote2019pacakge})  to generate new synthetic data for the minority class until the new training set is balanced. Then, we have a balanced data set in which each class is of size $n_1$. Following the default choice in \pkg{smotefamily}, we set the number of nearest neighbors $K=5$ in the oversampling process. 
	\item  Hybrid methods (Hybrid): we conduct a combination of random undersampling and SMOTE with the final training set consists of $n_h$ minority and majority observations with $n_h=\lfloor\sqrt{n_0*n_1}/n_0\rfloor*n_0$ where $\lfloor\cdot\rfloor$ is the floor function.
\end{itemize}

Regarding the base classification methods, we apply the following R packages or functions with their default parameters.
\begin{itemize}
	\item Logistic regression (\code{glm} function in base R).
	\item  Random forest (R Package \pkg{randomForest}, v4.6.14, \citealt{randomforest2002package}).
	\item Support vector machine (R Package \pkg{e1071}, v1.7.2, \citealt{svm2019package}).
	\item XGBoost (R Package \pkg{xgboost}, v0.90.0.2, \citealt{xgboost2019pacakge}).
\end{itemize}
Regarding the classification paradigms, some specifics are listed below.
\begin{itemize}
	\item  CS learning paradigm: we specify the cost $C_0=\mbox{IR}$ and $C_1 = 1$.
	\item NP paradigm: we use the NP umbrella algorithm as implemented in R package \pkg{nproc} v2.1.4, and set $\alpha=0.05$ and the tolerance level $\delta=0.05$. 
\end{itemize}

Denote by $|S|$ the cardinality of a set $S$.  Let $O=\{\text{CC, CS, NP}\}$, $T$=\{Original, Under, SMOTE, BLSMOTE, Hybrid\}, $C=\{\text{LR, RF, SVM, XGB}\}$ and $B=\{2^i,i=0,1,2,\ldots,7\}$.
Hence, there are $|O|\times|T|\times|C|\times|B|$ (480)  classification systems studied in this paper for a given imbalanced classification problem.

For each ensemble system, we evaluate the performance of different classifiers in terms of the following metrics reviewed in Section \ref{sec:eval_metric}: overall classification error (Risk), type I error, type II error, expected misclassification cost (Cost), $F$-score (class 0), and $F$-score (class 1). When the threshold varies for each classification method, we also report the area under ROC curve (ROC-AUC) and  the area under PR curve (PR-AUC (class 0) and PR-AUC (class 1)).

We repeat the experiment 100 times and report the average performance in terms of mean,
standard error,  and winning methods for each metric and classification paradigm combination. The results are summarized in Figures \ref{auc_ex1} to \ref{np_typeII_ex1_imb} as well as in Tables \ref{tb:ex1_1}, \ref{tb:ex1_2}, \ref{tb:ex2_1} and \ref{tb:ex2_2}. 

\subsection{Results and interpretations}\label{subsec::simu-results}
For each figure, we present the results of classification methods under each IR in the first four panels, while the last panel shows the optimal combination of resampling technique and base classification method under each IR.

Next, we provide some interpretations and insights from the figures and tables under each classification paradigm. 

\begin{figure}[!htbp]
	\centering
	\includegraphics[width=5 in]{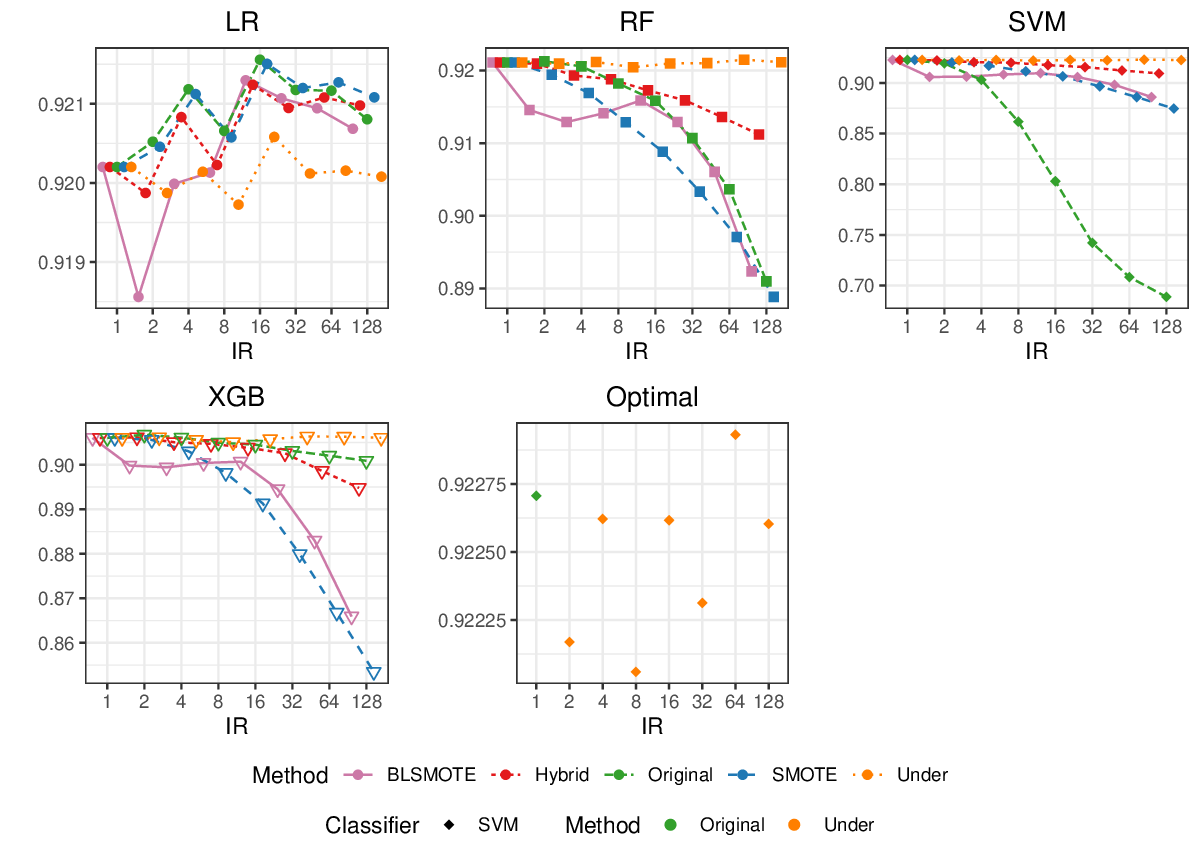}\\
	\caption{ROC-AUC of different methods in Example \ref{ex:1}(a). The minimum and maximum of  standard error: LR(0.0003, 0.0005), RF(0.0004,0.0007), SVM(0.0003,0.0029), XGB(0.0005, 0.0013).}\label{auc_ex1}
\end{figure}

\begin{figure}[!htbp]
	\centering
	\includegraphics[width=5 in]{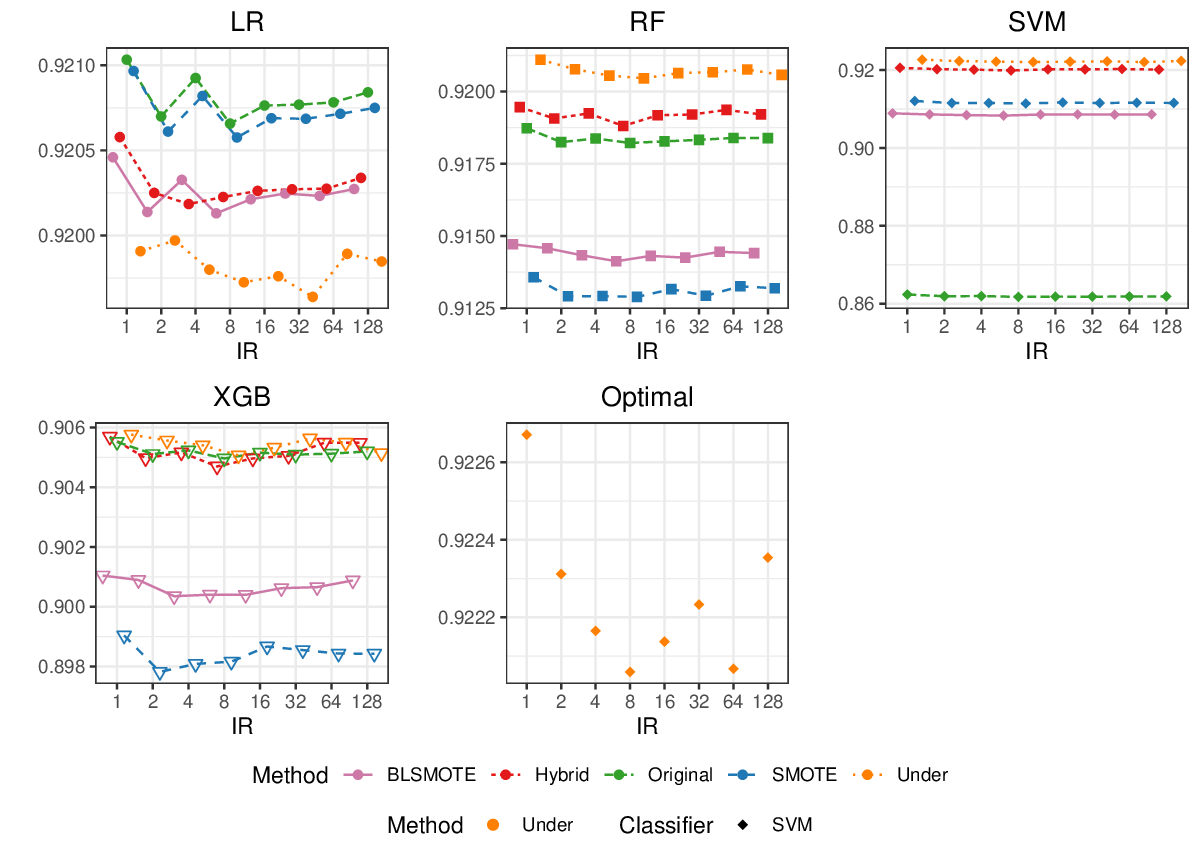}\\
	\caption{ROC-AUC of different methods in Example \ref{ex:1}(b). The minimum and maximum of  standard error: LR(0.0003, 0.0005), RF(0.0003,0.0006), SVM(0.0003,0.0010), XGB(0.0005, 0.0008).}\label{auc_ex1_imb}
\end{figure}

For Example \ref{ex:1}(a), where we vary the training and testing IR at the same time, we present the ROC-AUC in Figure \ref{auc_ex1} as an overall measure of classification methods without the need to specify the classification paradigm. First of all, LR is surprisingly stable for all resampling techniques across all IRs. Another study on the robustness of LR for imbalanced data can be found in \cite{owen2007infinitely}.  Then, from the panels corresponding to RF, SVM, and XGB, we suggest that it is essential to apply specific resampling techniques to keep the ROC-AUC at a high value when IR increases. 
For Example \ref{ex:1}(b) where we fix the training IR and vary the testing IR,  the ROC-AUC in Figure \ref{auc_ex1_imb} is more robust across the board.  
In addition, we report the range of the standard errors for each base classification method in the captions of Figures \ref{auc_ex1} and \ref{auc_ex1_imb}, and they are all very small. Thus, the standard error does not affect the determination of the optimal combination.
We omit the plots of ROC-AUC for Example  \ref{ex:2} as they look similar.

\subsubsection{Classical classification paradigm.} 
\begin{figure}[!htbp]
	\centering
	\includegraphics[width=5 in]{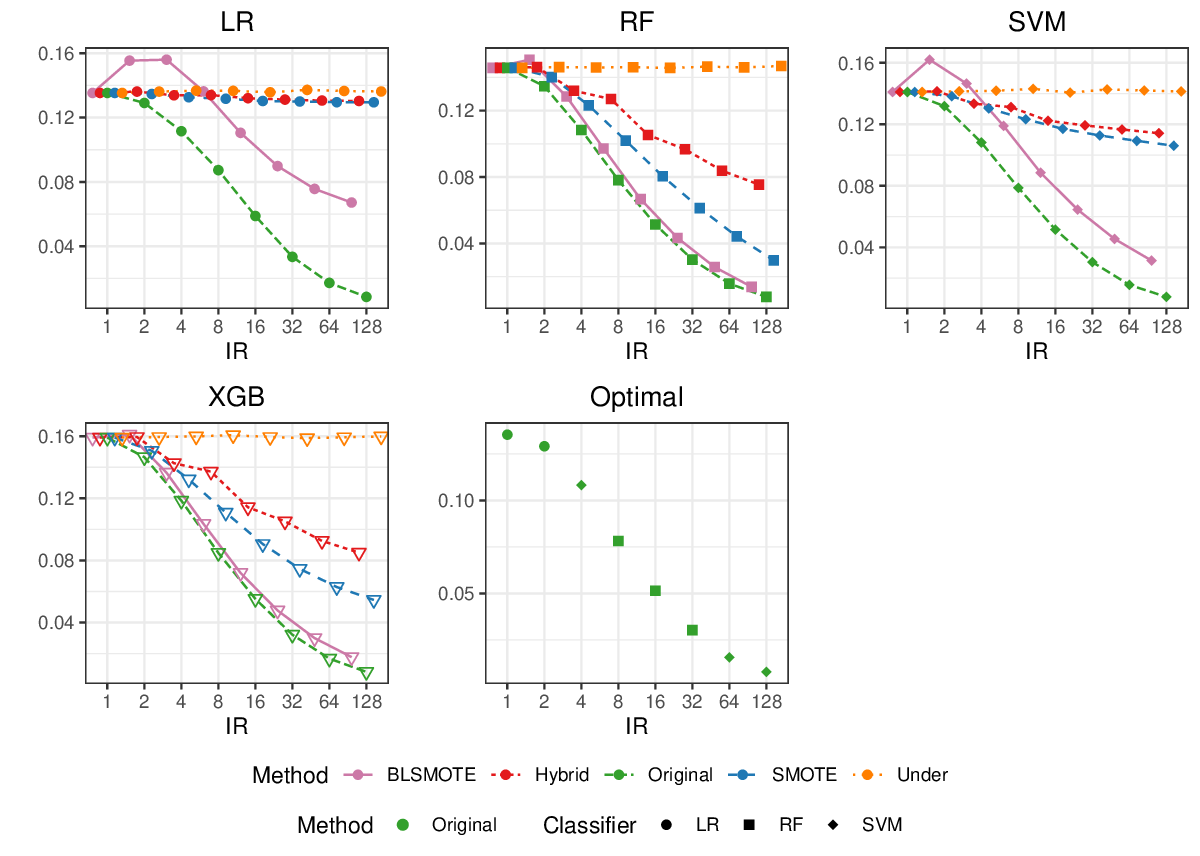}\\
	\caption{Risk of different methods under CC paradigm in Example \ref{ex:1}(a). The minimum and maximum of  standard error: LR(0, 0.0011), RF(0,0.0014), SVM(0,0.0012), XGB(0, 0.0014). }\label{cc_risk_ex1}
\end{figure}

\begin{figure}[!htbp]
	\centering
	\includegraphics[width=5 in]{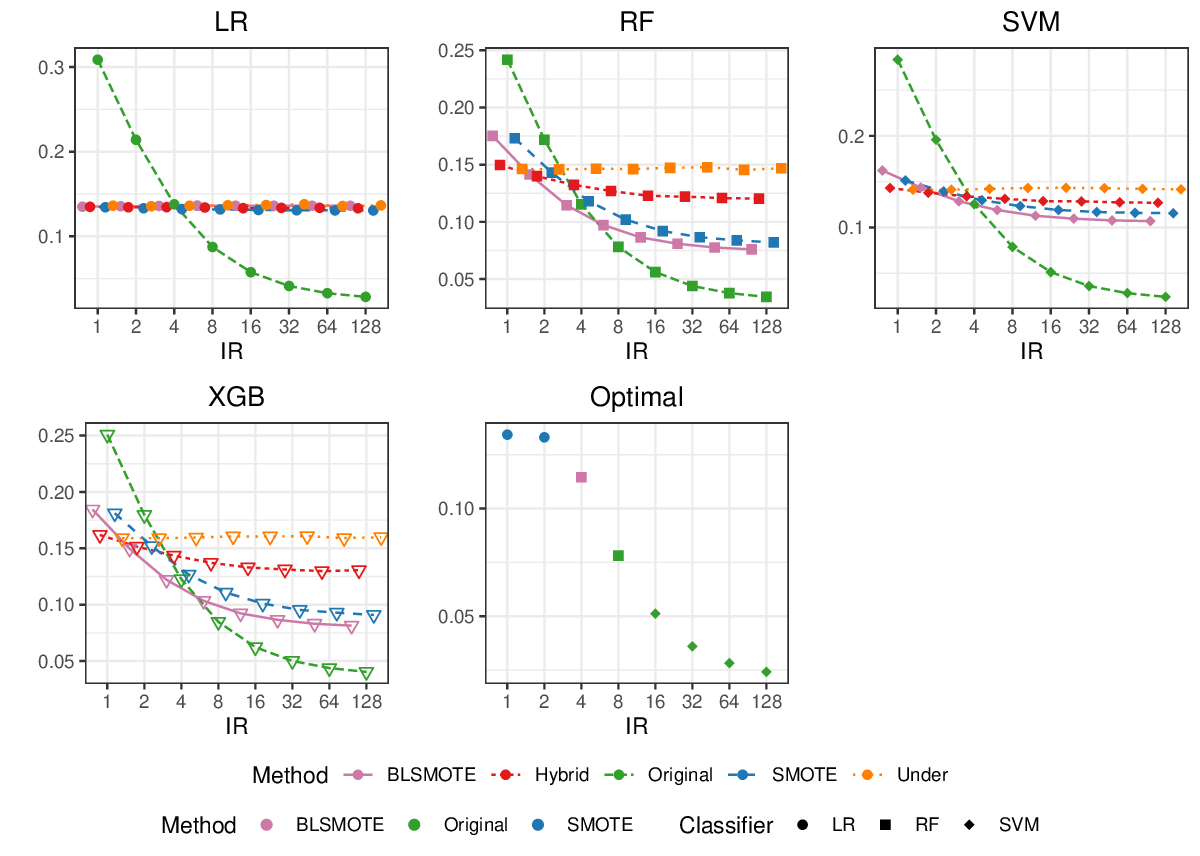}\\
	\caption{Risk of different methods under CC  paradigm in Example \ref{ex:1}(b). The minimum and maximum of  standard error: LR(0.0001, 0.0018), RF(0.0002,0.0016), SVM(0.0001,0.0016), XGB(0.0002, 0.0016).}\label{cc_risk_ex1_imb}
\end{figure}

We first focus on analyzing the results for Example \ref{ex:1}. 
Figures \ref{cc_risk_ex1} and \ref{cc_risk_ex1_imb} exhibit the risk of different methods. We observe that the empirical risk of all classifiers without resampling is smaller than that with any resampling technique in most cases, and decreases as IR increases. This is in line with our intuition that if the risk is the primary measure of interest, we would be better off not applying any resampling techniques. In addition, we observe that only undersampling leads to a stable risk when the IR increases for all four base classification methods considered.   Finally, the resampling techniques can make risk more stable across all IRs in Figure \ref{cc_risk_ex1_imb}.

\begin{figure}[!htbp]
	\centering
	\includegraphics[width=5 in]{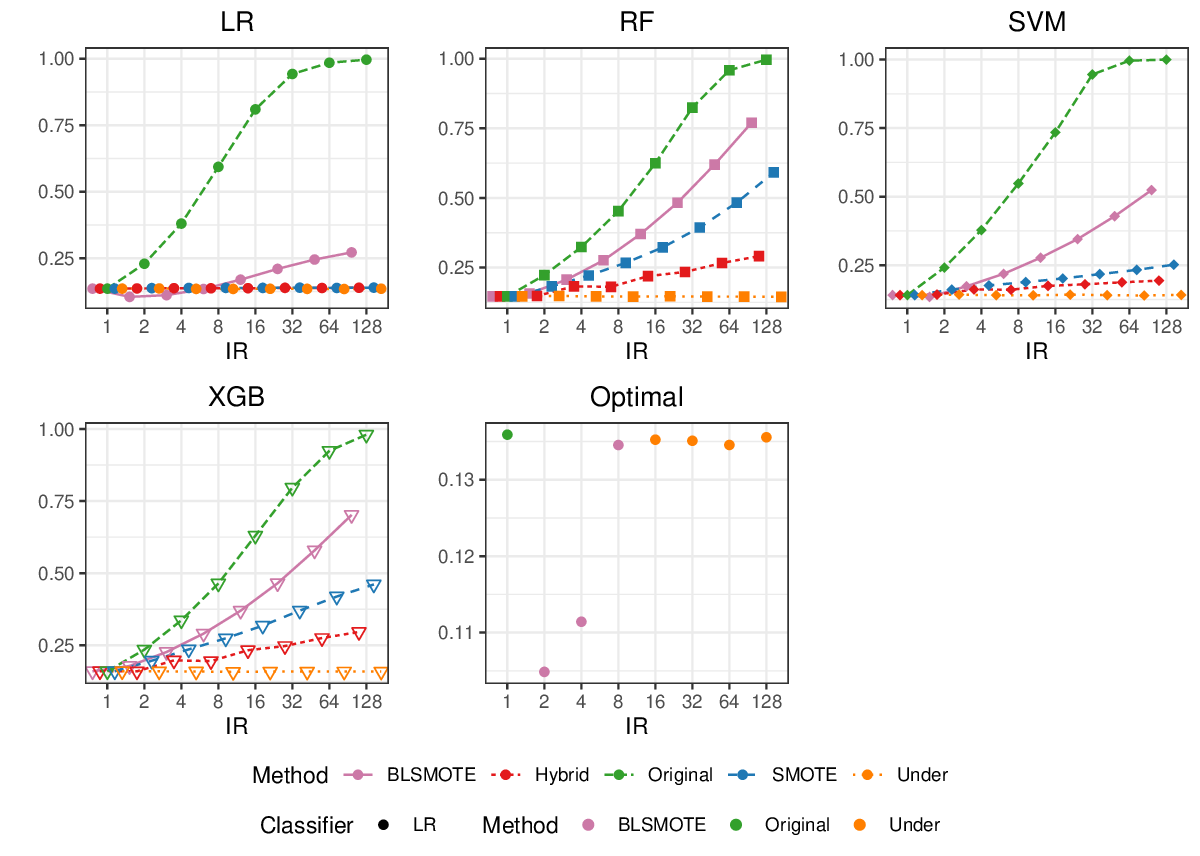}\\
	\caption{Type I error of different methods under CC paradigm  in Example \ref{ex:1}(a). The minimum and maximum of  standard error: LR(0, 0.0037), RF(0,0.0032), SVM(0,0.0034), XGB(0, 0.0027).}\label{cc_typeI_ex1}
\end{figure}

As mentioned in Section \ref{sec::class_paradigm}, minimizing the risk with imbalanced data could lead to large type I errors, demonstrated clearly in Figure \ref{cc_typeI_ex1}. By using the resampling techniques, however, we can have much better control over type I error as IR increases. In particular,  undersampling works well for all four classification methods. Lastly, we note that the optimal choices when IR $>1$ all involve resampling techniques. 

The figures for Example \ref{ex:2} convey a similar message as in Example \ref{ex:1} that we do not need any resampling if the goal is to minimize the risk. On the other hand, applying certain resampling techniques is critical to bring down the type I error and increase the ROC-AUC value. Again, we omit these figures to save space.

\subsubsection{Cost-Sensitive learning paradigm.} 
\begin{figure}[!htbp]
	\centering
	\includegraphics[width=5 in]{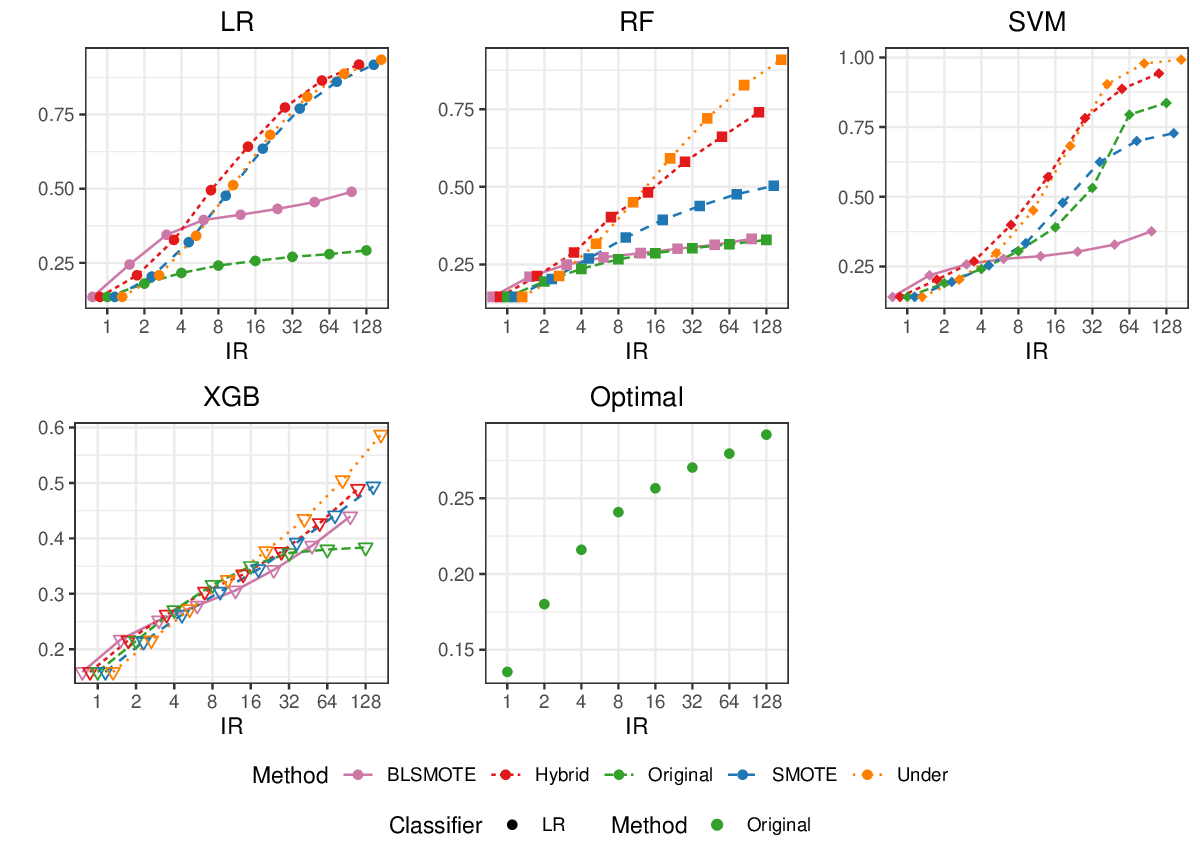}\\
	\caption{Cost of different methods under CS learning paradigm in Example \ref{ex:1}(a). The minimum and maximum of  standard error: LR(0.0006, 0.0066), RF(0.0007,0.0068), SVM(0.0004,0.0113), XGB(0.0004, 0.0066). }\label{cs_cost_ex1}
\end{figure}

\begin{figure}[!htbp]
	\centering
	\includegraphics[width=5 in]{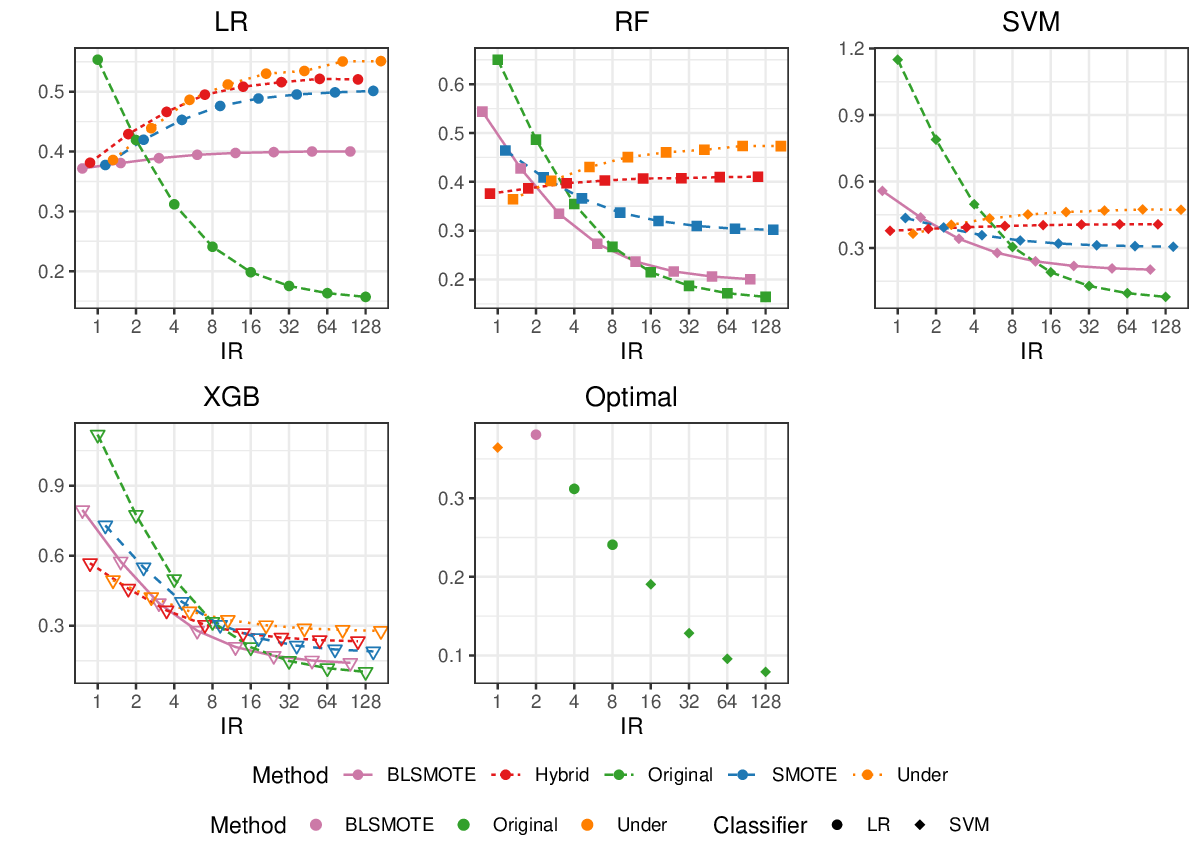}\\
	\caption{Cost of different methods under CS learning paradigm in Example \ref{ex:1}(b). The minimum and maximum of  standard error: LR(0.0006, 0.0061), RF(0.0008,0.0054), SVM(0.0003,0.0075), XGB(0.0005, 0.0065). }\label{cs_cost_ex1_imb}
\end{figure}

When we are in the CS learning paradigm, the objective is to minimize the expected total misclassification cost. We again first look at the results from Example \ref{ex:1}. Naturally, we would like to see the impact of the resampling techniques on different classification methods in terms of empirical cost, which is summarized in Figures \ref{cs_cost_ex1} and \ref{cs_cost_ex1_imb}. From the figures, we observe that no resampling leads to the smallest cost  in most cases. When IR is large, BLSMOTE leads to the smallest cost for SVM. 

\begin{figure}[!htbp]
	\centering
	\includegraphics[width=5 in]{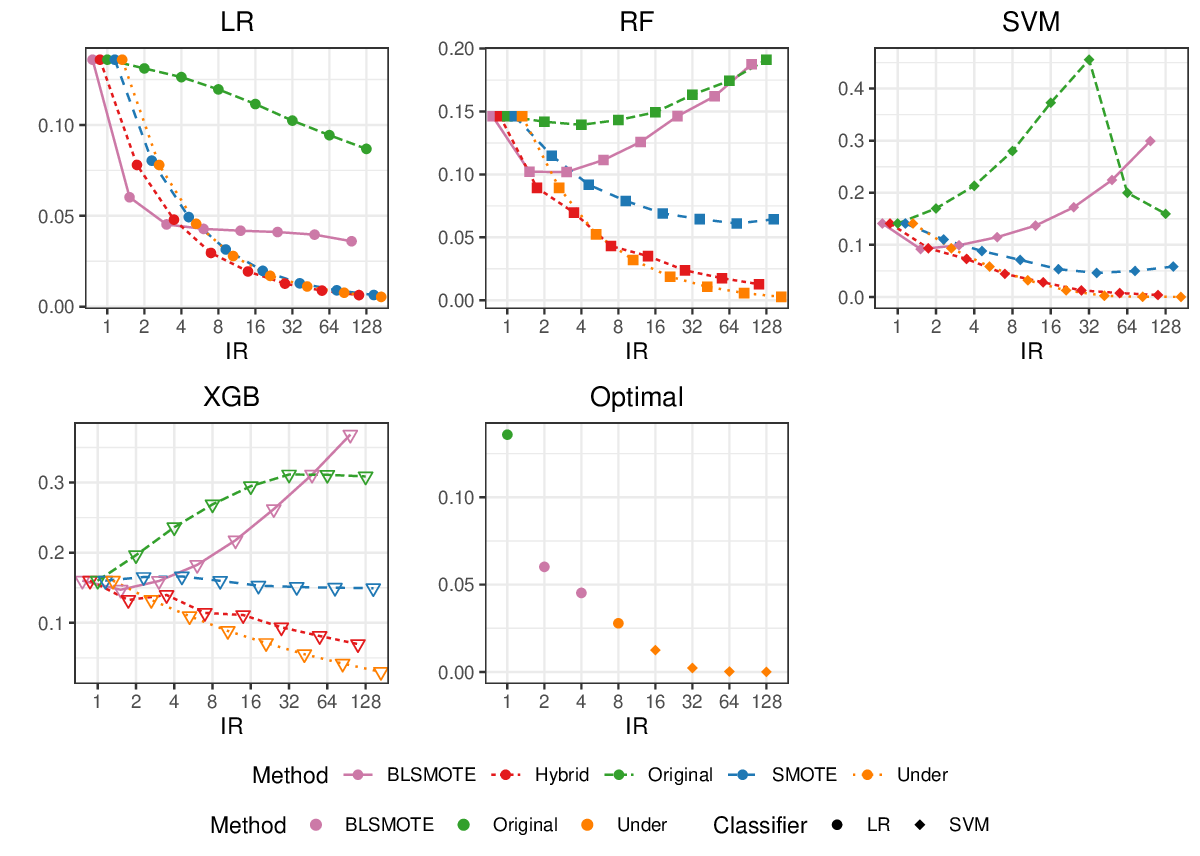}\\
	\caption{Type I error of different methods under CS learning paradigm in Example \ref{ex:1}(a). The minimum and maximum of  standard error: LR(0.0002, 0.0014), RF(0.0002,0.0017), SVM(0,0.0078), XGB(0.0007, 0.0019). }\label{cs_typeI_ex1}
\end{figure}

\begin{figure}[!htbp]
	\centering
	\includegraphics[width=5 in]{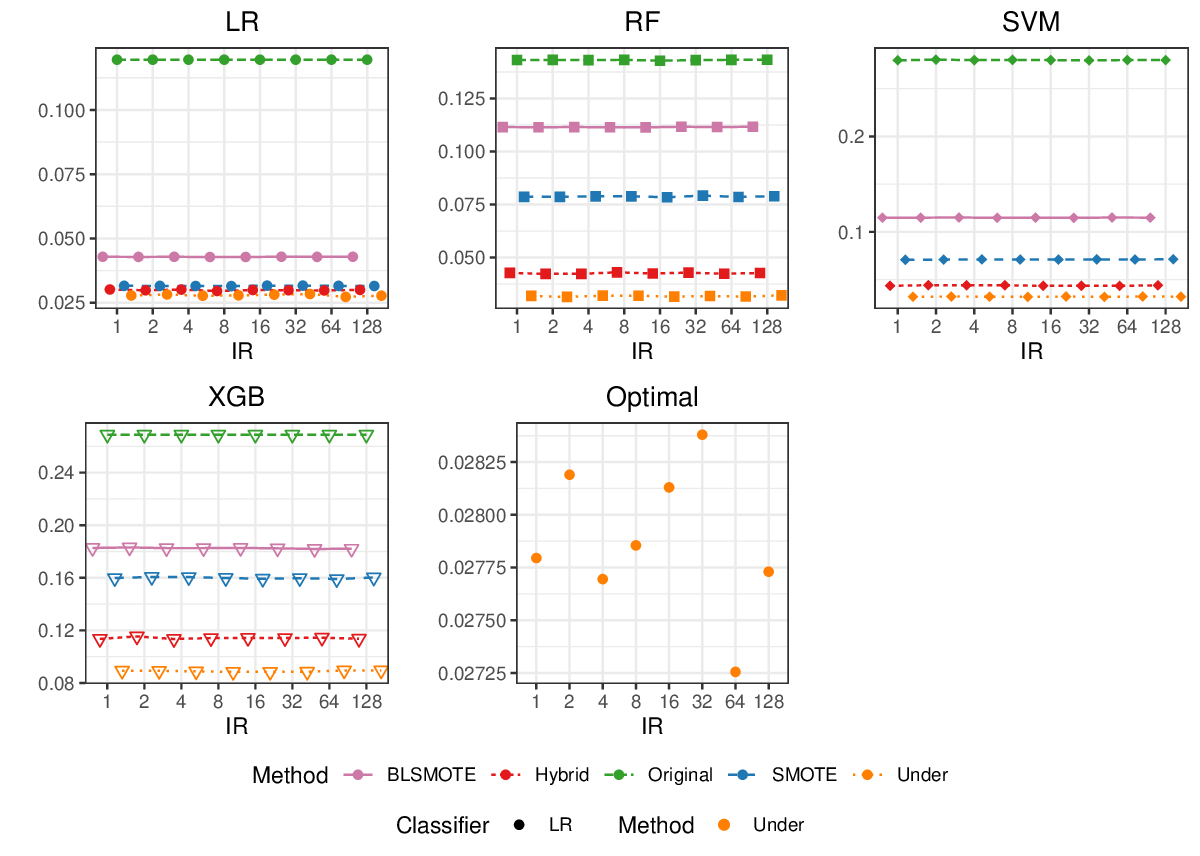}\\
	\caption{Type I error of different methods under CS learning paradigm in Example \ref{ex:1}(b). The minimum and maximum of  standard error: LR(0.0006, 0.0010), RF(0.0006,0.0015), SVM(0.0008,0.0019), XGB(0.0010, 0.0017). }\label{cs_typeI_ex1_imb}
\end{figure}

Now, we look at the results for type I error in Figures \ref{cs_typeI_ex1} and \ref{cs_typeI_ex1_imb}, where we discover that all classification methods benefit significantly from resampling techniques with undersampling being the best choice for most scenarios.

\subsubsection{Neyman-Pearson paradigm.}
\begin{figure}[!htbp]
	\centering
	\includegraphics[width=5 in]{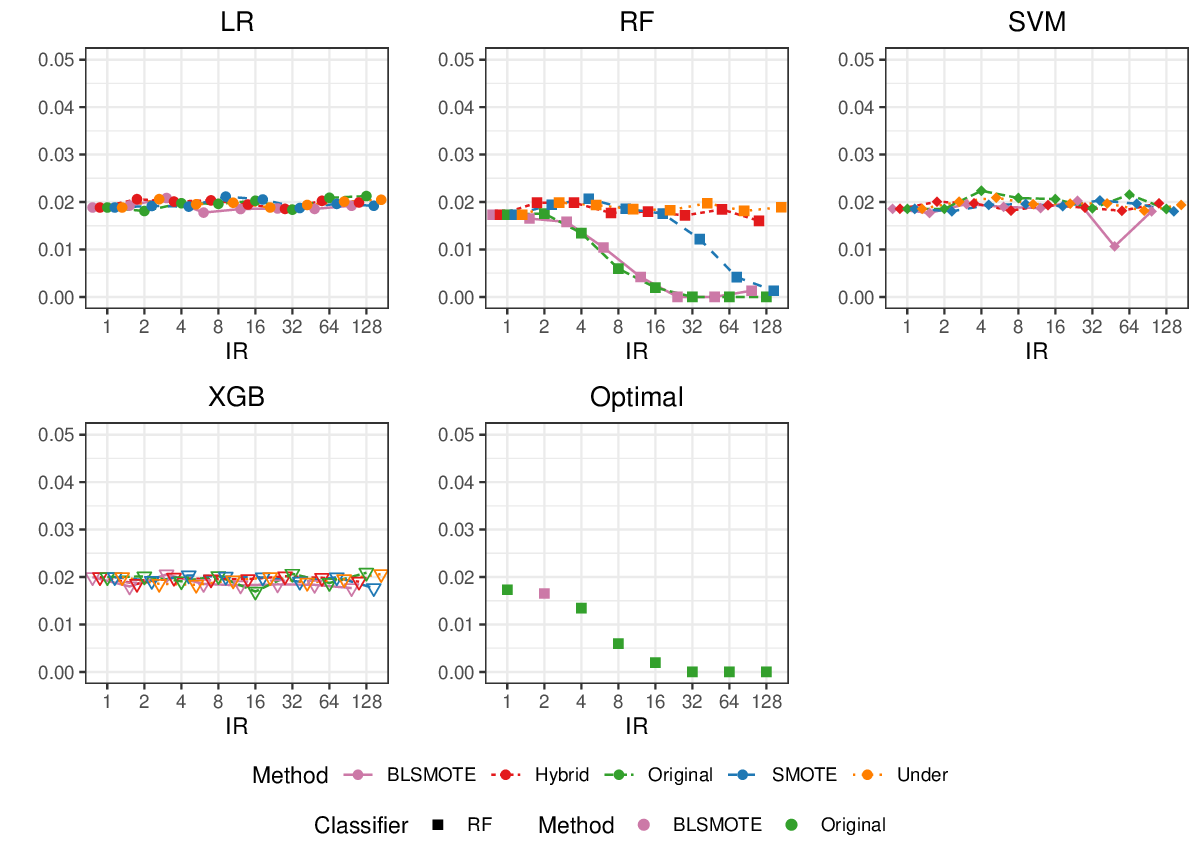}\\
	\caption{ Type I error of different methods under NP paradigm in Example \ref{ex:1}(a). The minimum and maximum of  standard error: LR(0.0010, 0.0014), RF(0,0.0015), SVM(0.0009,0.0015), XGB(0.0009, 0.0013).}\label{np_typeI_ex1}
\end{figure}

\begin{figure}[!htbp]
	\centering
	\includegraphics[width=5 in]{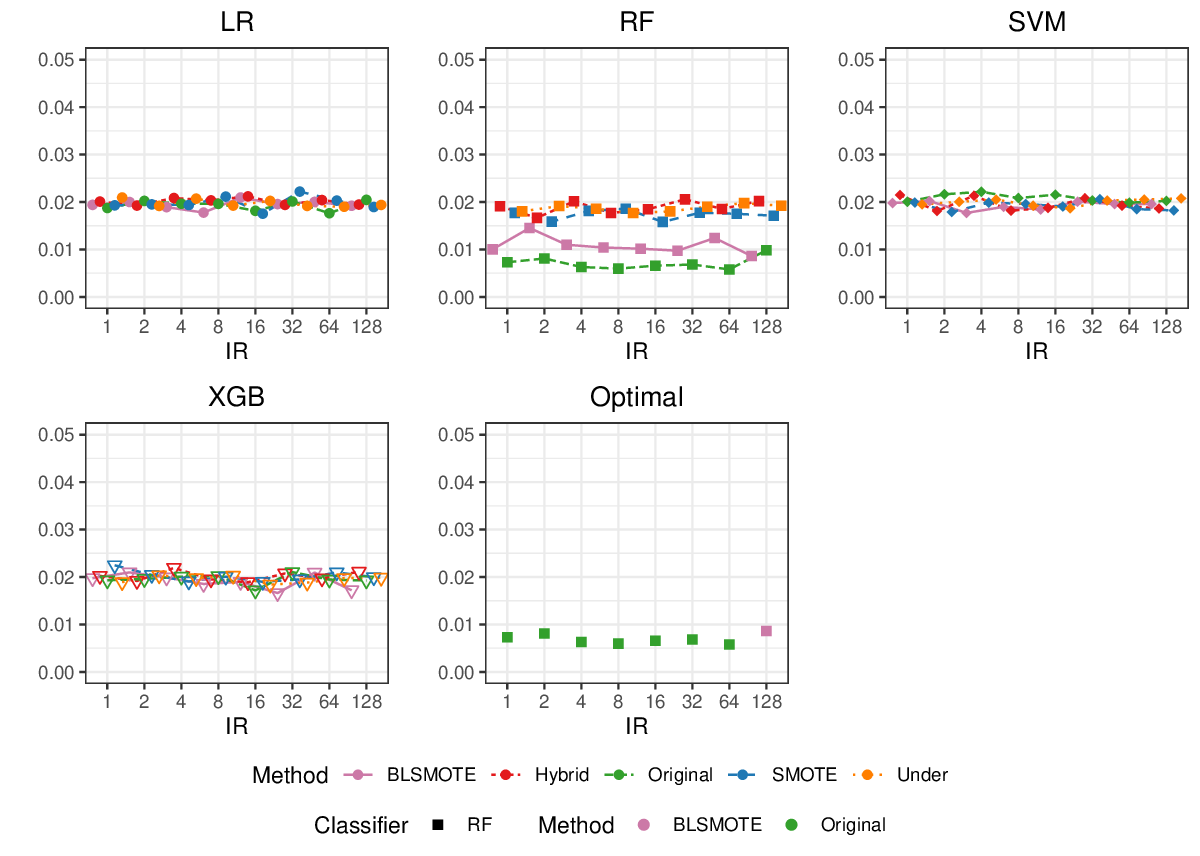}\\
	\caption{Type I error of different methods under NP paradigm in Example \ref{ex:1}(b). The minimum and maximum of  standard error: LR(0.0009, 0.0014), RF(0.0010,0.0017), SVM(0.0009,0.0015), XGB(0.0009, 0.0014). }\label{np_typeI_ex1_imb}
\end{figure}

The NP paradigm aims to minimize type II error while controlling type I error under a target level $\alpha$. In the current implementation, we set $\alpha=0.05$. From Figures \ref{np_typeI_ex1} and \ref{np_typeI_ex1_imb}, we observe that the type I errors are well-controlled under $\alpha$ throughout all IRs for all base classification methods in  Examples \ref{ex:1}(a) and \ref{ex:1}(b).

\begin{figure}[!htbp]
	\centering
	\includegraphics[width=5 in]{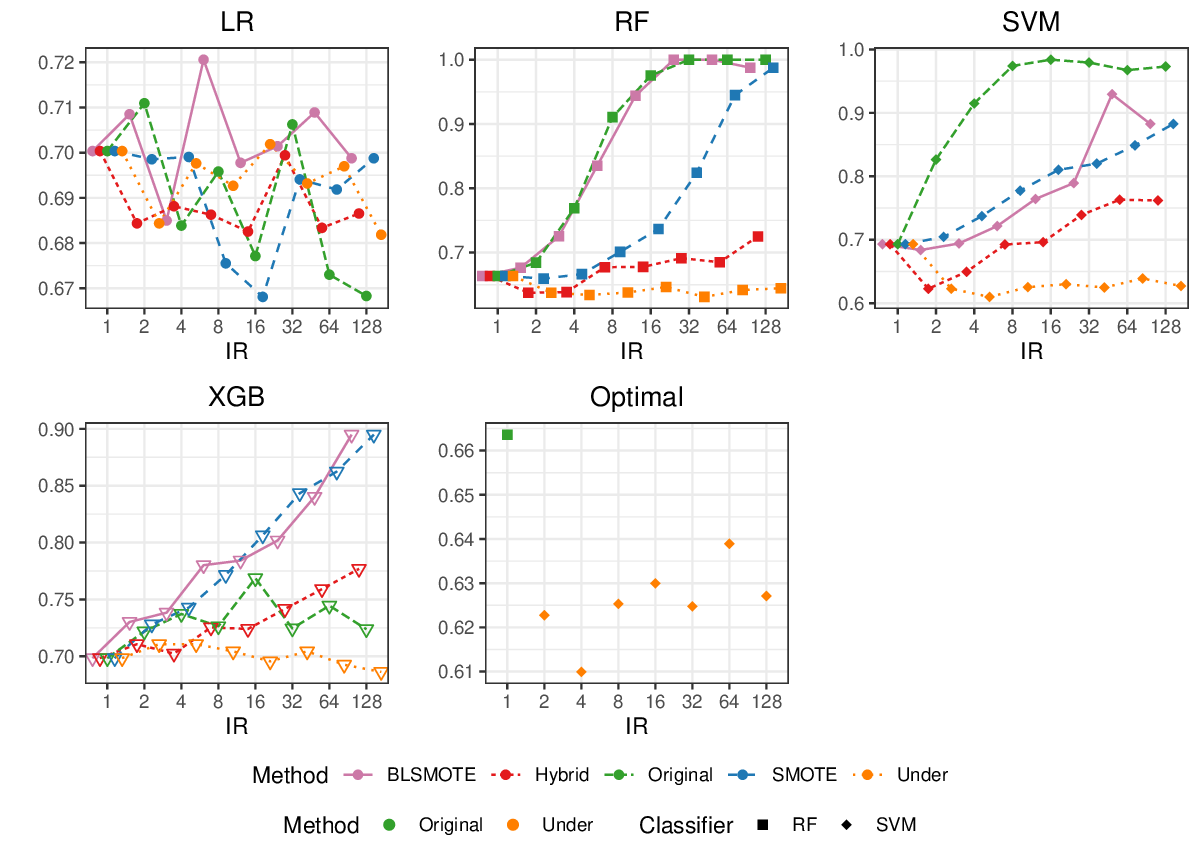}\\
	\caption{ Type II error of different methods under NP paradigm in Example \ref{ex:1}(a). The minimum and maximum of  standard error: LR(0.0142, 0.0182), RF(0,0.0219), SVM(0.0015,0.0149), XGB(0.0081, 0.0140).}\label{np_typeII_ex1}
\end{figure}

When we look at Figure \ref{np_typeII_ex1},  the benefits that resampling techniques can bring are apparent in most cases. Undersampling or hybrid resampling leads to a type II error well under control.  Moreover, Type II error is more robust when different IRs are selected for the test data set.

\begin{figure}[!htbp]
	\centering
	\includegraphics[width=5 in]{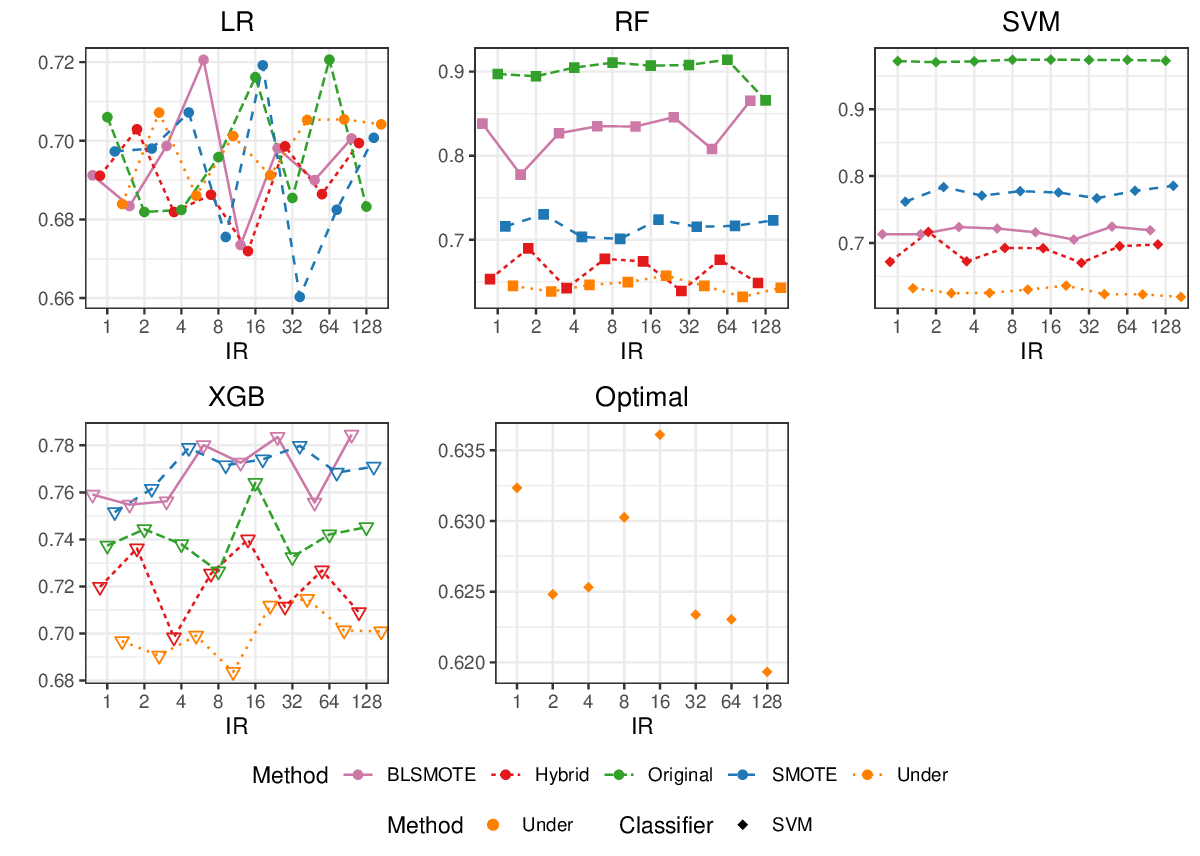}\\
	\caption{Type II error of different methods under NP paradigm in Example \ref{ex:1}(b). The minimum and maximum of  standard error: LR(0.0137, 0.0174), RF(0.0126,0.0232), SVM(0.0023,0.0153), XGB(0.0097, 0.0136).}\label{np_typeII_ex1_imb}
\end{figure}


For Example \ref{ex:2}, we have the same conclusion that  resampling techniques can help to reduce type II error with the type I error well-controlled under $\alpha$. 

\subsubsection{Summary.}
In addition to the plots, we summarize in Tables \ref{tb:ex1_1}, \ref{tb:ex1_2}, \ref{tb:ex2_1}, \ref{tb:ex2_2}  the  winning frequency  of resampling techniques and classification methods in terms of each evaluation metric of all IRs in Examples \ref{ex:1}(a), \ref{ex:1}(b), \ref{ex:2}(a), and \ref{ex:2}(b), respectively. 
The number in each cell of tables represents the winning frequency for each base classification method or each resampling technique for the given metric.   The numbers in bold  represent the most frequent winning combination of resampling techniques and classification methods.  Clearly, the optimal choices differ for different evaluation metrics, IRs, and data generation mechanisms. 
From these tables and the above figures, we can draw the following conclusions: 
\begin{enumerate}
	\item [(a)] All the classifiers can control the type I error under a certain level $\alpha$ under the  NP paradigm (see Figures \ref{np_typeI_ex1} and \ref{np_typeI_ex1_imb}).
	\item [(b)] For most base classification methods, ROC-AUC can usually benefit from  resampling techniques, whether or not the test class proportion is at the same level of imbalance  as the training set (see Figures \ref{auc_ex1} and \ref{auc_ex1_imb}).
	\item [(c)] Resampling techniques, in general, bring down the type I error  regardless of the classification paradigm (see Figures \ref{cc_typeI_ex1} and \ref{cs_typeI_ex1_imb}).
	\item [(d)]The optimal combination of base classification method and resampling technique should be interpreted together with both the paradigm and evaluation metric. For example, in Table \ref{tb:ex1_1},  the combination ``LR+Under'' leads to the minimal type I error under the CC paradigm.
	\item [(e)] When the training class proportion is fixed and IR varies for the test data set, the results are robust in most cases (see Figures \ref{cc_risk_ex1_imb}, \ref{cs_typeI_ex1_imb}, and \ref{np_typeII_ex1_imb}).
\end{enumerate}

	\begin{table}[!htbp]%
		\caption{ The frequency of winning methods  in Example 1(a).\label{tb:ex1_1}}
		\centering
		\begin{tabular}{ccccc|cccccc}
			\hline
			\diagbox[width= 6 em,trim=l]{Metric}{Method}    &   LR    & RF    &   SVM  &  XGB & BLSMOTE & Hybrid & Original & SMOTE & Under   \\
			\hline
			AUC                   &      0        &0     &  {\bf 8}   &  0  &0   &0   &1  &0  & {\bf 7}   \\
			PR-AUC(0)          &      0        &{\bf 5}     &  3   &  0  &1   &0   &{\bf 6}  &0  &1   \\ 
			PR-AUC(1)          &      0        &0     &  {\bf 8}   &  0  &0   &0   &1  &0  &{\bf 7}   \\
			\hline\hline
			CC-Type I           &     {\bf 8}        &   0   & 0    &  0  &3   &0   &1  &0  &{\bf 4}   \\
			CS-Type I           &     4        &   0   & {\bf 4}    &  0  &2   &0   &1  &0  &{\bf 5}   \\
			NP-Type I           &     0        &   {\bf 8}   & 0    &  0  &1   &0   &{\bf 7}  &0  &0   \\
			\hline
			Type I                &     12       &  8   & 4    &  0  &6   &0   &9  &0  &9   \\
			\hline\hline
			CC-Type II          &     1        &   0   & {\bf 7}    &  0  &0   &0   &{\bf 8}  &0  &0   \\	
			CS-Type II          &     1        &   0   & {\bf 4}    &  3  &1   &0   &{\bf 7}  &0  &0   \\	
			NP-Type II          &     0        &   1   & {\bf 7}    &  0  &0   &0   &1  &0  &{\bf 7}   \\
			\hline
			Type II               &     2        &   1   &18   &  3  &1    &0   &16 &0 &7\\
			\hline\hline
			CS-Cost             &     {\bf 8}        &   0   &0     &  0  &0    &0   &{\bf 8}   &0 &0 \\	
			\hline\hline
			CC-Risk             &     2        &   {\bf3}   &{\bf3}     &  0  &0    &0   &{\bf 8}   &0 &0 \\	
			NP-Risk             &     0        &   1   &{\bf 7}     &  0  &0    &0   &1   &0 &{\bf 7} \\	
			\hline
			Risk                  &     2        &   4   &10    &  0  &0    &0   &9   &0 &7 \\	
			\hline\hline
			CC-$F$-score(0)  &    3       &   {\bf 4}   &0     &  1  &{\bf 5}    &0   &1   &2 &0 \\	
			CS-$F$-score(0)  &    2       &   0   &{\bf 4}     &  2  &1    &0   &{\bf 7}   &0 &0 \\	
			NP-$F$-score(0)  &    0       &   2   &{\bf 6}     &  0  &0    &0   &1   &0 &{\bf 7} \\	
			\hline
			$F$-score(0)       &    5       &   6   &10    &  3  &6    &0   &9   &2 &7 \\
			\hline\hline
			CC-$F$-score(1)  &    2       &   1   &{\bf 5}    &  0  &0    &0   &{\bf 8}   &0 &0 \\	
			CS-$F$-score(1)  &    1       &   0   &{\bf 4}     &  3  &1    &0   &{\bf 7}   &0 &0 \\	
			NP-$F$-score(1)  &    0       &   1   &{\bf 7}     &  0  &0    &0   &1   &0 &{\bf 7}\\	
			\hline
			$F$-score(1)       &    3       &   2   &16    &  3  &1    &0   &16   &0 &7 \\
			\hline\hline
			Total                  &32         &   26  &77   &9    &15   &0   &75  &2  &52\\ 
			\hline
		\end{tabular}
	\end{table}

	\begin{table}[!htbp]%
		\caption{The frequency of winning methods  in Example 1(b).}\label{tb:ex1_2}
		\centering
		\begin{tabular}{ccccc|cccccc}
			\hline
			\diagbox[width= 6 em,trim=l]{Metric}{Method}    &   LR    & RF    &   SVM  &  XGB & BLSMOTE & Hybrid & Original & SMOTE & Under   \\
			\hline 
			AUC                   &      0        &0     &  {\bf 8}   &  0  &0   &0   &0  &0  &{\bf 8}   \\
			PR-AUC(0)          &      0        &{\bf 5}     &  3   &  0  &0   &0   &{\bf 8}  &0  &0   \\ 
			PR-AUC(1)          &      0        &0     &  {\bf 8}   &  0  &0   &0   &0  &0  &{\bf 8}   \\
			\hline\hline
			CC-Type I           &     {\bf 8}        &   0   & 0    &  0  &{\bf 7}   &0   &0  &0  &1   \\
			CS-Type I           &     {\bf 8}        &   0   & 0    &  0  &0   &0   &0  &0  &{\bf 8}   \\
			NP-Type I           &     0        &   {\bf 8}   & 0    &  0  &1   &0   &{\bf 7}  &0  &0   \\
			\hline
			Type I                &     16       &  8   & 0    &  0  &8   &0   &7  &0  &9   \\
			\hline\hline
			CC-Type II          &     2        &   2   & {\bf 4}    &  0  &1   &0   &{\bf 5}  &2  &0   \\	
			CS-Type II          &     0        &   0   & {\bf 8}    &  0  &0   &0   &{\bf 8}  &0  &0   \\	
			NP-Type II          &     0        &   0   & {\bf 8}    &  0  &0   &0   &0  &0  &{\bf 8}   \\
			\hline
			Type II               &     2        &   2   &20   &  0  &1    &0   &13 &2 &8\\
			\hline\hline
			CS-Cost             &     3        &   0   &{\bf 5}     &  0  &1    &0   &{\bf 6}   &0 &1 \\	
			\hline\hline
			CC-Risk             &     2        &   2   &{\bf 4}     &  0  &1    &0   &{\bf 5}   &2 &0 \\	
			NP-Risk             &     0        &   0   &{\bf 8}     &  0  &0    &0   &0   &0 &{\bf 8} \\	
			\hline
			Risk                  &     2        &   2   &12    &  0  &1    &0   &5   &2 &8 \\	
			\hline\hline
			CC-$F$-score(0)  &    3       &   2   &{\bf 3}     &  0  &2    &0   &{\bf 4}   &2 &0 \\	
			CS-$F$-score(0)  &    2       &   0   &{\bf 6}     &  0  &0    &0   &{\bf 8}   &0 &0 \\	
			NP-$F$-score(0)  &    0       &   0   &{\bf 8}     &  0  &0    &0   &0   &0 &{\bf 8} \\	
			\hline
			$F$-score(0)       &    5       &   2   &17    &  0  &2    &0   &12   &2 &8 \\
			\hline\hline
			CC-$F$-score(1)  &    1       &   2   &{\bf 5}     &  0  &1    &0   &{\bf 6}   &1 &0 \\	
			CS-$F$-score(1)  &    1       &   0   &{\bf 7}     &  0  &0    &0   &{\bf 8}   &0 &0 \\	
			NP-$F$-score(1)  &    0       &   0   &{\bf 8}     &  0  &0    &0   &0   &0 &{\bf 8} \\	
			\hline
			$F$-score(1)       &    2       &   2   &20    &  0  &1    &0   &14   &1 &8 \\
			\hline\hline
			Total                  &30         &   21  &93   &0    &14   &0   &65  &7  &58\\ 
			\hline
		\end{tabular}
	\end{table}

	\begin{table}[!htbp]%
		\caption{The frequency of winning methods  in Example 2(a).}\label{tb:ex2_1}
		\centering
		\begin{tabular}{ccccc|cccccc}
			\hline
			\diagbox[width= 6 em,trim=l]{Metric}{Method}    &   LR    & RF    &   SVM  &  XGB & BLSMOTE & Hybrid & Original & SMOTE & Under   \\
			\hline 
			AUC                   &      0        &0     &  {\bf 8}   &  0  &0   &0   &1  &0  &{\bf 7}   \\
			PR-AUC(0)          &      0        &0     &  {\bf 8}   &  0  &0   &0   &1  &0  &{\bf 7}   \\ 
			PR-AUC(1)          &      0        &0     &  {\bf 8}   &  0  &0   &0   &1  &0  &{\bf 7}   \\
			\hline\hline
			CC-Type I           &     0        &   0   & {\bf 8}    &  0  &0   &0   &1  &0  &{\bf 7}   \\
			CS-Type I           &     {\bf 6.1} &   0   & 1.9    &  0  &0.9   &1.4   &1  & {\bf 2.4}  &2.3   \\
			NP-Type I           &     3        &  {\bf 5}    & 0    &  0  &2   &0   &{\bf 5}  &0  &1   \\
			\hline
			Type I                &     9.1       &  5   & 9.9    &  0  &2.9   &1.4   &7  &2.4  &10.3   \\
			\hline\hline
			CC-Type II          &    {\bf 7}        &   0   & 1    &  0  &0   &0   &{\bf 8}  &0  &0   \\	
			CS-Type II          &     0        &   0   & 3    &  {\bf 5}  &1   &0   &{\bf 7}  &0  &0   \\	
			NP-Type II          &     0        &   5   & {\bf 3}    &  0  &0   &2   &1  &0  &{\bf 5}   \\
			\hline
			Type II               &     7        &   5   &7   &  5  &1    &2   &16 &0 &5\\
			\hline\hline
			CS-Cost             &     0        &   {\bf 3}   &{\bf 4}     &  1  &{\bf 6}    &0   &1   &1 &0 \\	
			\hline\hline
			CC-Risk             &     {\bf 7}        &   0   &1     &  0  &0    &0   &{\bf 8}   &0 &0 \\	
			NP-Risk             &     0        &   4  & {\bf 4}      &  0  &0    &2   &1   &0 &{\bf 5} \\	
			\hline
			Risk                  &     7        &   4   &5    &  0  &0    &2   &9   &0 &5 \\	
			\hline\hline
			CC-$F$-score(0)  &    0       &   0   &{\bf 8}     &  0  &0    &0   &1   &0 &{\bf 7} \\	
			CS-$F$-score(0)  &    0       &   0   &{\bf 7}     &  1  &{\bf 5}    &0   &1   &1 &1 \\	
			NP-$F$-score(0)  &    0       &   5   &{\bf 3}     &  0  &0    &2   &1   &0 &{\bf 5} \\	
			\hline
			$F$-score(0)       &    0       &   5   &18    &  1  &5    &2   &3   &1 &13 \\
			\hline\hline
			CC-$F$-score(1)  &   {\bf 7}        &   0   &1     &  0  &0    &0   &{\bf 8}   &0 &0 \\	
			CS-$F$-score(1)  &    0       &   0   &3     &  {\bf 5}  &1    &0   &{\bf 7}   &0 &0 \\	
			NP-$F$-score(1)  &    0       &   5   &{\bf 3}     &  0  &0    &2   &1   &0 &{\bf 5} \\	
			\hline
			$F$-score(1)       &    7       &   5   &7    &  5  &1    &2   &16   &0 &5 \\
			\hline\hline
			Total                  &30.1         &   27  &74.9   &12    &15.9   &9.4   &55  &4.4  &59.3\\ 
			\hline
		\end{tabular}
	\end{table}

	\begin{table}[!ht]%
		\caption{The frequency of winning methods  in Example 2(b).}\label{tb:ex2_2}
		\centering
		\begin{tabular}{ccccc|cccccc}
			\hline
			\diagbox[width= 6 em,trim=l]{Metric}{Method}    &   LR    & RF    &   SVM  &  XGB & BLSMOTE & Hybrid & Original & SMOTE & Under   \\
			\hline 
			AUC                   &      0        &0     &  {\bf 8}   &  0  &0   &0   &0  &0  &{\bf 8}   \\
			PR-AUC(0)          &      0        &0     &  {\bf 8}   &  0  &0   &0   &0  &0  &{\bf 8}   \\ 
			PR-AUC(1)          &      0        &0     &  {\bf 8}   &  0  &0   &0   &0  &0  &{\bf 8}   \\
			\hline\hline
			CC-Type I           &     0        &   0   & {\bf 8}    &  0  &0   &0   &0  &0  &{\bf 8}   \\
			CS-Type I           &     {\bf 8}        &   0   &0    &  0  &{\bf 2}   &{\bf 2}   &0  & {\bf 2}  &{\bf 2}  \\
			NP-Type I           &     0        &  {\bf 8}    & 0    &  0  &2   &0   &{\bf 6}  &0  &0   \\
			\hline
			Type I                &     8      &  8   & 8    &  0  &4   &2   &6  &2  &10   \\
			\hline\hline
			CC-Type II          &    {\bf 8}        &   0   & 0    &  0  &0   &0   &{\bf 8}  &0  &0   \\	
			CS-Type II          &     0        &   0   & 0    &  {\bf 8}  &0   &0   &{\bf 8}  &0  &0   \\	
			NP-Type II          &     0        &   {\bf 6}   &  2    &  0  &0   &{\bf 3}   &0  &0  &{\bf 5}   \\
			\hline
			Type II               &     8        &   6   &2   &  8  &0    &3   &16 &0 &5\\
			\hline\hline
			CS-Cost             &     1        &   1    & 2     &  {\bf 4}  &1   &0   &{\bf 4}   &1 &2 \\	
			\hline\hline
			CC-Risk             &     {\bf 7}        &   0   &1     &  0  &0    &0   &{\bf 7}   &0 &1 \\	
			NP-Risk             &     0        &   {\bf 6}  & 2      &  0  &0    &{\bf 4}   &0   &0 &4 \\	
			\hline
			Risk                  &     7        &   6  &3    &  0  &0    &4   &7   &0 &5 \\	
			\hline\hline
			CC-$F$-score(0)  &    0       &   0   &{\bf 8}     &  0  &3    &1  &0  &0 &{\bf 4} \\	
			CS-$F$-score(0)  &    0       &   1   &{\bf 4}     &  3  &{\bf 4}    &0   &3   &1 &0 \\	
			NP-$F$-score(0)  &    0       &   {\bf 6}   &2     &  0  &0    &{\bf 4}   &0   &0 & 4 \\	
			\hline
			$F$-score(0)       &    0       &   7   &14    &  3  &7    &5   &3   &1 &8 \\
			\hline\hline
			CC-$F$-score(1)  &   {\bf 8}        &   0   &0     &  0  &0    &0   &{\bf 8}   &0 &0 \\	
			CS-$F$-score(1)  &    0       &   0   &0     &  {\bf 8}  &0    &0   &{\bf 8}   &0 &0 \\	
			NP-$F$-score(1)  &    0       &   {\bf 6}   & 2     &  0  &0    &{\bf 3}   &0   &0 & {\bf 5} \\	
			\hline
			$F$-score(1)       &    8       &   6   &2    &  8  &0    &3   &16   &0 &5 \\
			\hline\hline
			Total                  &32         &   34  &55   &23    &12   &17   &52  &4  &59\\ 
			\hline
		\end{tabular}
	\end{table}

\newpage
\section{Real Data-Credit Card Fraud Detection}\label{sec:real}
The Credit Card Transaction Data is available at \url{http://kaggle.com/mlg-ulb/creditcardfraud}. It includes credit card transactions made  in September 2013 by European cardholders. In particular, it contains transactions that occurred in two days, where we have 492 frauds out of 284,807 transactions. Therefore, this data set is highly imbalanced with an imbalance ratio (IR) about 578 (284,315/492).  Due to confidentiality issues,  the website does not provide the original features and more background information about this data set.  Features V1, V2, $\ldots$, V28 are the principal components obtained with PCA.  The only features which have not been transformed with PCA are ``Time'' and ``Amount''.  Feature ``Time'' contains the seconds elapsed between each transaction and the first transaction in the data set.   The feature ``Amount'' is the transaction amount. They are scaled to zero mean and unit variance.  Using the feature ``Class'', we redefine ``0'' as the fraud class (class 0) and ``1'' as the no-fraud class (class 1). We use the features V1, V2, $\ldots$, V28, Time and Amount as predictor variables for the classification methods.

 We specify the imbalance ratio (IR) as 128 for training data set  and extract a subsample from this large dataset. In particular, we randomly sample $n_0=300$ data points from class 0 (fraud) and  $n1=n0*\mbox{IR}=38,400$ from class 1 (no-fraud). This procedure creates our training data set. The test data set contains a random sample of $m_0=192$ for class 0 and $m_1=m_0*\mbox{IR}_{test}$ for class 1 from the remaining data, where  $\mbox{IR}_{test}$ varies in $\{2^i, i=0,1,\cdots,7\}$. This splitting mechanism implies that IR will be different for the training and test data sets. 

\begin{figure}[!htbp]
 	\centering
 	\includegraphics[width=5 in]{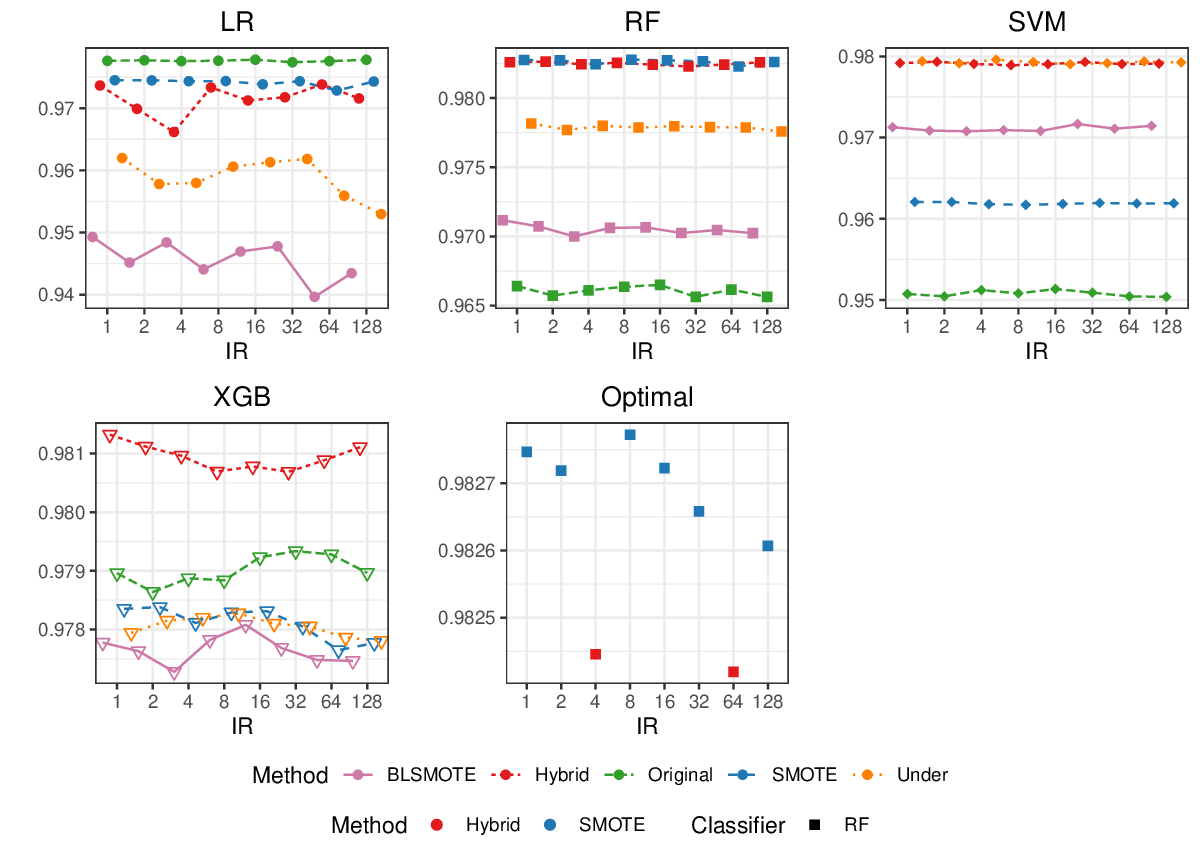}\\
 	\caption{ ROC-AUC of different methods  in real data. The minimum and maximum of  standard error: LR(0.0005, 0.0089), RF(0.0005,0.0009), SVM(0.0005,0.0014), XGB(0.0005, 0.0007).}\label{auc_credit}
 \end{figure}

\begin{figure}[!htbp]
 	\centering
 	\includegraphics[width=5 in]{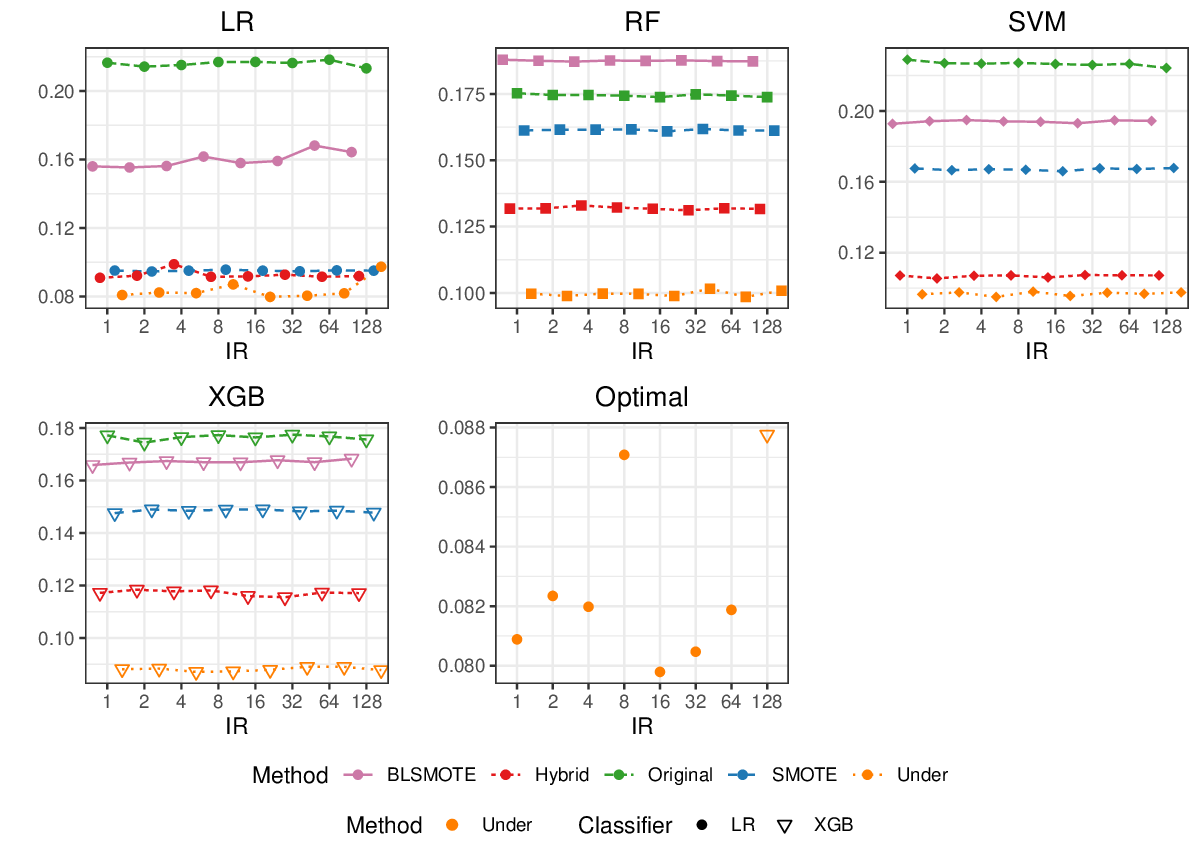}\\
 	\caption{ Type I error of different methods under CC paradigm in real data. The minimum and maximum of  standard error: LR(0.0020, 0.0086), RF(0.0019,0.0024), SVM(0.0019,0.0036), XGB(0.0018, 0.0023).}\label{cc_credit_typeI}
 \end{figure}

The remaining implementation details are the same as in Section \ref{subsec:imple_details}. We still repeat the experiment 100 times and report the average performance and frequency of winning methods by the mean for each metric and classification paradigm combination. The frequency of winning methods were summarized in Table \ref{tb:credit_1} and report Figures \ref{auc_credit} and \ref{cc_credit_typeI} and omit the other figures since they convey similar information to that in Section \ref{subsec::simu-results}.

 From Figures \ref{auc_credit} and \ref{cc_credit_typeI},  resampling techniques are in general beneficial for the metrics in most cases. In addition, most of the results are robust when the test IR increases. This is  consistent with the simulation results. Table \ref{tb:credit_1} shows that the combination ``RF+Hybrid'' has the top performance. Note that this appears to be different from the choices implied by Tables \ref{tb:ex1_1}-\ref{tb:ex2_2}, which again show that the best performing method highly depends on the data generation process. This actually agrees with our understanding of SVM vs. RF in that RF may be more effective than SVM in a more complex scenario.  Moreover, the optimal methods depend on the learning paradigm and evaluation metrics. For example, if our objective is to minimize the overall risk under the CC paradigm, ``RF+SMOTE'' is the best choice in Table \ref{tb:credit_1};  if our objective is to minimize the type II error while controlling the type I error under a specific level,  ``RF+Hybrid''  performs the best. Therefore, there is no universal best combination for the imbalanced classification problem.

	\begin{table}[!ht]%
		\caption{The frequency of winning methods  when  IR of test data varies in credit fraud detection.}\label{tb:credit_1}
		\centering
		\begin{tabular}{ccccc|cccccc}
			\hline
			\diagbox[width= 6 em,trim=l]{Metric}{Method}    &   LR    & RF    &   SVM  &  XGB & BLSMOTE & Hybrid & Original & SMOTE & Under   \\
			\hline 
			AUC                   &      0        &{\bf 8}     &  0   &  0  &0   &2   &0  &{\bf 6}  &0   \\
			PR-AUC(0)          &      0        &{\bf 8}     &   0   &  0  &0   &0   &0  &{\bf 8}  &0  \\ 
			PR-AUC(1)          &      0        &{\bf 8}      &  0   &  0  &0   &{\bf 8}   &0  &0  & 0   \\
			\hline\hline
			CC-Type I           &     {\bf 7}        &   0   & 0    &  1  &0   &0   &0  &0  &{\bf 8}   \\
			CS-Type I           &    0        &   0   &{\bf 8}    &  0  &0   & 0   &0  &0  & {\bf 8}  \\
			NP-Type I           &     0        &  {\bf 8}    & 0    &  0  &0   &0   &{\bf 8}  &0  &0   \\
			\hline
			Type I                &     7      &  8   & 8    &  1  &0   &0   &8  &0  &16   \\
			\hline\hline
			CC-Type II          &    0        &   0   & {\bf 4}    & {\bf 4}   &0   &0   &{\bf 8}  &0  &0   \\	
			CS-Type II          &     0        &   0   & 0    &  {\bf 8}  &0   &0   &{\bf 8}  &0  &0   \\	
			NP-Type II          &     0        &   {\bf 6}   &  2    &  0  &0   &{\bf 6}   &0  &0  & 2   \\
			\hline
			Type II               &     0        &   6   &6   &  12  &0    &6   &16 &0 &2\\
			\hline\hline
			CS-Cost             &     0        &   1    & 0     &  {\bf 7}  &0   &2   &{\bf 3}   &3 &0 \\	
			\hline\hline
			CC-Risk             &     1        &  {\bf 3}    &0     &  4  &0    &3   &1   &{\bf 4} &0 \\	
			NP-Risk             &     0        &   {\bf 6}  & 2      &  0  &0    &{\bf 6}   &0   &0 &2 \\	
			\hline
			Risk                  &     1        &   9  &2    &  4  &0    &9   &1   &4 &2 \\	
			\hline\hline
			CC-$F$-score(0)  &    1       &   {\bf 3}   &0     &  {\bf 4}  &0    &{\bf 4}  &1  &{\bf 3} &0 \\	
			CS-$F$-score(0)  &    0       &   0   &0     &  {\bf 8}  & 0    &0   &{\bf 6}   &2 &0 \\	
			NP-$F$-score(0)  &    0       &   {\bf 8}   &0     &  0  &0    &{\bf 8}   &0   &0 & 0 \\	
			\hline
			$F$-score(0)       &    1       &   11   &0    &  12  &0    &12   &7   &5 &0 \\
			\hline\hline
			CC-$F$-score(1)  &   1        &   {\bf 4}   &0     &  3  &0    &3   &1  &{\bf 4} &0 \\	
			CS-$F$-score(1)  &    0       &   0   &0     &  {\bf 8}  &0    &0   &{\bf 6}   &2 &0 \\	
			NP-$F$-score(1)  &    0       &   {\bf 6}   & 2     &  0  &0    &{\bf 6}   &0   &0 &  2 \\	
			\hline
			$F$-score(1)       &    1       &   10   &2    &  11  &0    &9   &7   &6 &2 \\
			\hline\hline
			Total                  &10         &   69  &18   &47    &0   &48   &42  &32  &22\\ 
			\hline
		\end{tabular}
	\end{table}

\newpage
\section{Discussion}\label{sec:discussion}
In this paper, we review the imbalanced classification with a paradigm-based view. 
In addition to the few take-away messages we offered in the simulation section, the main message from the review is that there is no single best approach to imbalanced classification. The optimal choice for resampling techniques and base classification methods highly depends on the classification paradigms, evaluation metric, as well as the severity of imbalancedness (imbalance ratio).

Admittedly, we only considered a selective list of resampling techniques and base classification methods. There are many other combinations that are worth further consideration. In addition, we presented results from two simulated data generation processes as well as a real data set, which could be unrepresentative for specific applications. We suggest practitioners adapt our analysis process for evaluating different choices for imbalanced classification to align with their data generation mechanism.  

Furthermore,  in our numerical experiments,  all base classification methods were applied  using the corresponding R-packages with their default parameters. Note that although we didn't tune the parameters due to the already-extensive simulation settings, it is well known that parameter tuning could further improve the performance of a classifier in certain situation. For example, the parameter $k$ in  SMOTE \cite{chawla2002smote} can be selected via cross-validation. 
We leave a systematic study of the impact of parameter tuning on imbalanced classification as a future research topic. 

Lastly, we focused on binary classification throughout the review. We expect similar interpretations and conclusions from multi-class imbalanced classification.

\newpage
\bibliographystyle{plainnat}
\bibliography{references}

\end{document}